\documentclass[journal]{IEEEtran}
\usepackage[colorlinks, citecolor=blue]{hyperref}

\usepackage{cite}
\usepackage{xcolor}
\usepackage{url,hyperref,booktabs}
\usepackage{siunitx}
\usepackage{arydshln}
\usepackage{pifont}

\ifCLASSINFOpdf
  \usepackage[pdftex]{graphicx}
  \DeclareGraphicsExtensions{.pdf,.jpeg,.png}
  \usepackage{epstopdf}
  \usepackage{subcaption}
\else
  \usepackage[dvips]{graphicx}
\fi
\usepackage{amsmath}
\usepackage{amssymb}
\usepackage{multirow}
\usepackage[switch]{lineno}
\usepackage{array}

\usepackage{stfloats}
\usepackage[symbol]{footmisc}
\usepackage{soul}

\makeatletter
\def\endthebibliography{%
  \def\@noitemerr{\@latex@warning{Empty `thebibliography' environment}}%
  \endlist
}
\makeatother

\begin{document}
%
\title{\textcolor{black}{Selective State Space Model for Monaural Speech Enhancement}}
%
%
%

\author{Moran Chen,~
        Qiquan Zhang, \IEEEmembership{Member,~IEEE,}
        Mingjiang Wang,~
        Xiangyu Zhang,~
        Hexin Liu,~\IEEEmembership{Member,~IEEE,}
        Eliathamby Ambikairaiah,~\IEEEmembership{Senior Member,~IEEE,}
        and Deying Chen
\thanks{Moran Chen, Qiquan Zhang, Mingjiang Wang, and Deying Chen are with the School of Electronic and Information Engineering, Harbin Institute of Technology, Shenzhen, 518055 China (e-mail: {moranchen1121}@gmail.com; {zhang.qiquan}@outlook.com; {mjwang}@hit.edu.cn; {dychen}@hit.edu.cn).}
\thanks{Eliathamby Ambikairaiah is with the School of Electrical Engineering and Telecommunications, The University of New South Wales, Sydney, 2052, Australia~(e-mail: {e.ambikairajah}@unsw.edu.au).}
\thanks{Qiquan Zhang is also with the School of Electrical Engineering and Telecommunications, University of New South Wales, Sydney, Australia.
}
\thanks{Xiangyu Zhang is with the School of Electrical Engineering and Telecommunications, University of New South Wales, Sydney, Australia.}
\thanks{Hexin Liu is with the College of Computing and Data Science, Nanyang Technological University, Singapore.}


}

%
%

\markboth{Journal of \LaTeX\ Class Files,~Vol.~14, No.~8, August~2019}%
{Shell \MakeLowercase{\textit{et al.}}: Bare Demo of IEEEtran.cls for IEEE Journals}
%



\maketitle
\begin{abstract}



\textcolor{black}{Voice user interfaces (VUIs) have facilitated the efficient interactions between humans and machines through spoken commands. Since real-word acoustic scenes are complex, speech enhancement plays a critical role for robust VUI. Transformer and its variants, such as Conformer, have demonstrated cutting-edge results in speech enhancement. However, both of them suffers from the quadratic computational complexity with respect to the sequence length, which hampers their ability to handle long sequences. Recently a novel State Space Model called Mamba, which shows strong capability to handle long sequences with linear complexity, offers a solution to address this challenge. In this paper, we propose a novel hybrid convolution-Mamba backbone, denoted as MambaDC, for speech enhancement. Our MambaDC marries the benefits of convolutional networks to model the local interactions and Mamba's ability for modeling long-range global dependencies. We conduct comprehensive experiments within both basic and state-of-the-art (SoTA) speech enhancement frameworks, on two commonly used training targets. The results demonstrate that MambaDC outperforms Transformer, Conformer, and the standard Mamba across all training targets. Built upon the current advanced framework, the use of MambaDC backbone showcases superior results compared to existing \textcolor{black}{SoTA} systems. This sets the stage for efficient long-range global modeling in speech enhancement.}

\end{abstract}

\begin{IEEEkeywords}
\textcolor{black}{Speech enhancement, Mamba, selective state space model, consumer voice user interface}
\end{IEEEkeywords}

%
\IEEEpeerreviewmaketitle

\section{Introduction}\label{sec:1}
%
%
%
%


\IEEEPARstart{T}{he} \textcolor{black}{landscape of human-machine interaction in consumer electronics~\cite{TCE-HMI-IoT} have undergone remarkable evolution over time. The industry has made continuous efforts to improve user experience, progressing from traditional interfaces reliant on manual inputs to more intuitive and seamless approaches like touchscreens and gestures. Voice User Interaction (VUI) ~\cite{VUI1, VUI2, VUI3}, in which users engage with devices using spoken commands, represents a significant leap in this evolution. This paradigm shift not only streamlines user interactions but also open up new possibilities for hands-free and natural communication with technology.}

\textcolor{black}{Nevertheless, in practical scenarios, the operational acoustic environments for consumer electronics are often complex. Speech signals are inevitably distorted by background noises during transmission, posing great challenges to robust VUI~\cite{chentce,VUI_AUTOMATION}. To mitigate this issue, it is critical to deploy a speech enhancement system to isolate the clean speech from the corrupted speech to improve speech perceptual quality and intelligibility. Traditional approaches often exploit the statistical properties of the speech and noise signals and derive a filter function~\cite{loizou,zhang2019,mmse2017}. However, these approaches are incapable of eliminating rapidly changing noise sources. Over the last ten years, with the advent of deep learning, speech enhancement has achieved considerable advancements~\cite{overview2018,DeepMMSE,demcus}. Since the self-attention mechanism effectively captures long-range global interactions, Transformers~\cite{attention2017} have facilitated recent advances in many speech processing tasks, such as automatic speech recognition and speech enhancement~\cite{zhang2024empirical,tgsa,mpsenet}. Despite the notable success, the computational complexity of self-attention scales quadratically with respect to the sequence length, which makes Transformers very resource-intensive.} 

\textcolor{black}{More recently, the Mamba architecture~\cite{gu2023mamba}, a novel selective State Space Model (SSM), has exhibited remarkable potential in long-sequence modeling. In contrast to Transformer, Mamba, which employs a selective mechanism with hardware-optimized implementation, operates with linear computational complexity. A number of Mamba-based backbones have been proposed for computer vision~\cite{liu2024vmambavisualstatespace,zhuvision,guo2024mambair} and natural language processing tasks~\cite{lieber2024jamba}, showcasing promising results. This has catalyzed its rapid adoption in speech and audio tasks~\cite{mambaspeech,miyazaki2024exploring,jiang2024speech,jiang2024dual}. In the most recent study~\cite{mambaspeech}, we substantiate that Mamba surpasses Transformer in speech enhancement, while offering faster training and inference speeds. Moreover, the advances of Conformer~\cite{conformer-asr}, a convolution-augmented Transformer, inspires us to explore the potential of exploiting convolution network to augment Mamba backbone. This paper \textcolor{black}{further} proposes MambaDC, a hybrid convolution-SSM architecture, for speech enhancement. MambaDC is capable of capturing both localized fine-grained feature patterns and long-range global contextual dependencies with linear computational complexity. We extensively evaluate our MambaDC within basic and state-of-the-art (SoTA) speech enhancement frameworks, across different training targets.}

\textcolor{black}{\textcolor{black}{In short, the main contributions of this study are summarized as follows:}
\begin{itemize}
  \setlength\itemsep{0.5pt}
  \item \textcolor{black}{We explore the potential of incorporating convolutional network to augment Mamba for speech enhancement.}
  \item \textcolor{black}{We introduce a simple yet effective hybrid convolution-Mamba architecture (MambaDC) to enhance the original Mamba’s ability to capture local fine-grained features.}
  \item \textcolor{black}{We demonstrate that our MambaDC consistently delivers substantially superior performance over Transformer, Conformer, and the vanilla Mamba, surpassing current SoTA baseline systems. This paves the way for future innovations in network designs that effectively capture both local and global dependencies for speech enhancement.}
\end{itemize}}

\textcolor{black}{The remainder of this paper is organized as follows. Section \ref{sec:2} describes the related work including speech enhancement and state space models. We formulates the research problem in Section \ref{sec:3}. In Section \ref{sec:4}, we detail our proposal, neural speech enhancement with selective state space model. Section \ref{sec:5} describes the experimental setup. The experimental results and analysis are presented in Section \ref{sec:6}. Finally, Section \ref{sec:7}  concludes this paper.}

\section{Related Work}\label{sec:2}
\subsection{\textcolor{black}{Speech Enhancement}}

\textcolor{black}{Traditionally, given the assumption on the statistical distributions (e.g., complex Gaussian distribution) about speech and noise, a filter function can be derived and then applied to noisy spectrum to obtain the estimate of clean speech. In the past decade, the use of deep learning techniques has enabled remarkable progress in speech enhancement over traditional schemes. Deep learning methods can be broadly grouped into generative schemes and predictive schemes.}

\textcolor{black}{Predictive schemes dominate the filed of speech enhancement, where a neural network is optimized to learn a mapping from the noisy speech to the clean speech. Common methods include waveform mapping, time-frequency (T-F) mapping and masking. The waveform mapping methods often optimize a neural network with an explicit encoder-decoder structure to directly reconstruct the clean waveform from the noisy waveform~\cite{2018raw,demcus,conformer-se}. In contrast, T-F domain methods operate in the T-F representation, such as the \textcolor{blue}{magnitude spectrum}~\cite{tfaj}, the log-power spectrum~\cite{yongxu2015}, and the complex spectrum~\cite{8910352}. T-F mapping and masking methods optimize a neural network to predict the clean T-F representation or a T-F mask given a noisy T-F representation, respectively. The most commonly used T-F mask include ideal ratio mask (IRM)~\cite{wang2013}, spectral magnitude mask (SMM)~\cite{wang2013}, phase-sensitive mask (PSM)~\cite{psm}, and complex ideal ration mask (cIRM)~\cite{cirm}. The estimate of clean T-F representation by applying the predicted mask to the noisy T-F representation. In this paper, IRM and PSM are used for evaluation.}

\textcolor{black}{Generative approaches follow a different paradigm, training a neural network model to learn a prior distribution over clean speech data~\cite{segmse,storm,richter2023speech}. They seeks to learn the inherent characteristics in speech, such as T-F structure. This prior knowledge can be utilized to infer clean speech from noisy speech input that may fall outside the learned distribution. Unlike predictive methods that output a single best estimate, generative models allow to generate multiple valid estimates. Generative methods include likelihood-based models for learning explicit speech distribution such as the variational autoencoder (VAE)~\cite{bando2018statistical,leglaive2018variance,fang2021variational}, the generative adversarial networks (GANs) for implicit distribution estimates~\cite{metricgan,metricgan+,SEGAN}, and more recent diffusion-based generative models~\cite{song2019generative,ho2020denoising,sohl2015deep}.}

\textcolor{black}{Many types of network architectures have been applied for speech enhancement. Among the earlier ones is the multi-layer perceptron (MLP), which fails to leverage long-range dynamic dependencies~\cite{yongxu2015}. With the ability to model the long-range context interactions, long-short term memory recurrent neural network (LSTM-RNN) showcase superior performance~\cite{2014lstm,2015lstm,chenlstm}, especially in the generalization ability to unseen speakers. Despite the superiority of LSTM-RNN models, the nature of the sequential modeling prohibits their use in many applications. Subsequently, a convolutional network, termed temporal convolutional network (TCN) has demonstrated comparable or better performance to LSTM-RNN, with significantly fewer trainable parameters and faster training speed~\cite{tcn_se,zhang2019monaural,tfa}. Specifically, TCN incorporates residual skip connection and 1-D dilated convolution, which allows for building a very large receptive field.}

\textcolor{black}{Transformers~\cite{attention2017} has demonstrated the latest advancements in a variety of speech processing tasks, such as speech enhancement~\cite{zhang2024exploration,zhang2023ripple}. The multi-head self-attention (MHSA) module, the core component in Transformers, computes the interactions between all the time frames in parallel, allowing the model to learn the global interactions efficiently. Furthermore, the reference~\cite{conformer-asr} proposes combining convolutional network and Transformer (Conformer) to effectively capture both local and global interactions, showcasing SoTA performance in ASR and speech enhancement~\cite{conformer-se}. More recently, a selective state space model named Mamba has shown great potential as an alternative to Transformer in speech enhancement~\cite{mambaspeech}.
}

\subsection{\textcolor{black}{State Space Models}}
\textcolor{black}{The State Space Model (SSM) based models~\cite{guefficiently} provide an alternative to Transformers for modeling long-range contextual dependencies. A selective SSM termed Mamba~\cite{gu2023mamba} has recently been introduced, featuring a data-dependent SSM layer and forming a versatile language model backbone. It surpasses Transformers at multiple scales on extensive real-world datasets and offers the advantage of linear scalability with respect to the sequence length. Mamba has recently been successfully applied in many domains, such as natural language processing~\cite{lieber2024jamba}, computer vision~\cite{zhuvision,liu2024vmambavisualstatespace}, and speech processing~\cite{mambaspeech,jiang2024speech,miyazaki2024exploring}. In particularly, vision Mamba (Vim)~\cite{zhuvision} utilizes bidirectional Mamba to learn dynamic global context and achieves impressive performance. VMamba~\cite{liu2024vmambavisualstatespace} introduces a cross-scan module (CSM) for a more global context modeling, in which a four-way selective scan is utilized to integrate information from all surrounding tokens.}

\textcolor{black}{In speech processing domains, the most recent study~\cite{mambaspeech} explores Mamba as an alternative to Transformer in speech enhancement and automatic speech recognition in both causal and non-causal configurations, and demonstrates the great potential to be next-generation backbone for speech processing. \textcolor{black}{In addition, Mamba architecture has demonstrated remarkable success in speech separation~\cite{jiang2024dual}, spoken language understanding, speech summarization~\cite{miyazaki2024exploring}, and self-supervised speech processing~\cite{zhang2024rethinking}.}
}

\begin{figure}[h]
\begin{center}
\centerline{\includegraphics[width=0.9\columnwidth]{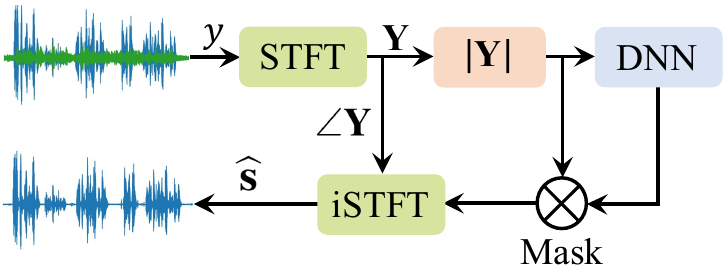}}
\caption{The overall diagram of a typical time-frequency neural speech enhancement system.}
\label{fig1}
\end{center}
\vskip -0.3in
\end{figure}

\section{\textcolor{black}{Problem Formulation}}\label{sec:3}
\subsection{Signal Model}\label{sec:3.1}
\textcolor{black}{Given the clean speech signal $\boldsymbol{s}\in \mathbb{R}^{1\times T}$ and the noise signal $\boldsymbol{d}\in \mathbb{R}^{1\times T}$, the noisy mixture signal $\boldsymbol{y}\in \mathbb{R}^{1\times T}$ can be modeled as:  
\begin{equation} \label{equ:b}
\boldsymbol{y}[t] = \boldsymbol{s}[t] + \boldsymbol{d}[t],
\end{equation}
where $t\!\in\!\{1,2,3,...,T\}$ denotes the index of discrete time samples. The time-frequency representation of the noisy mixture, \textcolor{black}{$\boldsymbol{y}$}, is extracted using the short-time Fourier transform (STFT):}
\textcolor{black}{\begin{equation} 
\mathbf{Y}_{l,k}=\mathbf{S}_{l,k}+\mathbf{D}_{l,k}
\end{equation}
\textcolor{black}{where $\mathbf{S}_{l,k}$, $\mathbf{D}_{l,k}$, and $\mathbf{Y}_{l,k}$}} represent the complex-valued STFT coefficients of the clean speech, noise, and noisy mixture, respectively, at the $k$-th frequency bin of the $l$-th time frame. 


\subsection{Training Targets}\label{sec:3.2}
\textcolor{black}{As shown in Figure~\ref{fig1}, a typical neural T-F speech enhancement solution takes the input the STFT magnitude spectrogram of the noisy speech \textcolor{black}{$\mathbf{Y}_{l,k}$} and optimizes a DNN to estimate a T-F mask \textcolor{black}{$\widehat{\mathbf{M}}_{l,k}$} to separate the clean speech:
\textcolor{black}{\begin{equation}
\widehat{\mathbf{S}}_{l,k}\!=\!\mathbf{Y}_{l,k}\cdot \widehat{\mathbf{M}}_{l,k}. 
\end{equation}}
\textcolor{black}{The enhanced waveform $\widehat{\mathbf{s}}$ of the speech signal is reconstructed from the predicted spectrum of clean speech \textcolor{black}{$\widehat{\mathbf{S}}_{l,k}$} via inverse STFT. Without loss of generality, we employ two commonly used T-F masks to conduct extensive performance evaluations, as summarized next.}}

\subsubsection{Ideal Ratio Mask}
\textcolor{black}{One of the most popular T-F masks is the \textcolor{black}{ideal ratio mask (IRM)}~\cite{targets}, which is defined as:
\textcolor{black}{\begin{equation}\label{IRM}
\text{IRM}_{l,k}=\left(\frac{|\mathbf{S}_{l,k}|^{2}}{|\mathbf{S}_{l,k}|^{2}+|\mathbf{D}_{l,k}|^{2}}\right)^{1/2}
\end{equation}}
\textcolor{black}{where $\left|\mathbf{S}_{l,k}\right|$ and $\left|\mathbf{D}_{l,k}\right|$} denote the STFT spectral magnitude of clean speech and noise, respectively. The IRM value falls within the ranges of 0 to 1.}

\subsubsection{\textcolor{black}{Phase-Sensitive Mask}}
\textcolor{black}{The phase-sensitive mask (PSM) involves both spectral magnitude and phase errors, enabling the predicted magnitude spectrum to compensate for the use of the noisy phase spectrum~\cite{psm}:
\textcolor{black}{\begin{equation}\label{SMM}
\text{PSM}_{l,k}=\frac{\left|\mathbf{S}_{l,k}\right|}{|\mathbf{Y}_{l,k}|}\cos(\phi_{\mathbf{s}} -\phi_{\mathbf{y}})
\end{equation}
where $\left|\mathbf{Y}_{l,k}\right|$} denotes spectral magnitude of the noisy speech. $\phi_{\mathbf{s}}$ and $\phi_{\mathbf{y}}$ respectively denote the phases of clean spectrum and noisy spectrum. The PSM value is clipped to $[0,1]$ to fit the output range of the sigmoidal activation function.}

\section{\textcolor{black}{Speech Enhancement With Selective State Space Model}}\label{sec:4}

\subsection{\textcolor{black}{Preliminaries}}
\textcolor{black}{\textbf{State Space Model.} State Space Models (SSMs) can be regarded as the continuous linear time-invariant (LTI) systems, which originates from the classic Kalman filter. It takes a time-dependent set of inputs $u(t)\!\in\!\mathbb{R}$ and maps it into a set of outputs \textcolor{black}{$z(t)\!\in\!\mathbb{R}$} through a hidden state $\mathbf{h}(t)\in\mathbb{R}^{N}$. Mathematically, the mapping process of SSMs can be formulated as a linear ordinary differential equation (ODE):
\textcolor{black}{
\begin{equation}\label{continue_equation}
\begin{aligned}
    \mathbf{h}'(t) &= \mathbf{A}\mathbf{h}(t) + \mathbf{B}u(t), \\
    z(t) &= \mathbf{C}\mathbf{h}(t) + \mathbf{D}u(t),
\end{aligned}
\end{equation}}
where $\mathbf{A}\!\in\!\mathbb{R}^{N\times N}$, $\mathbf{B}\!\in\!\mathbb{R}^{N\times 1}$, $\mathbf{C}\!\in\!\mathbb{R}^{1\times N}$, and $\mathbf{D}\!\in\!\mathbb{R}$ are state matrix, input matrix, output matrix, and feed-forward matrix.}

\textcolor{black}{\textbf{Discretization of SSM.} To integrate SSMs into practical deep learning architecture, we first apply the discretization process to continue-time SSMs in advance. To be specific, a timescale parameter $\mathbf{\Delta}$ is typically adopted to transform the continuous matrices $\mathbf{A}, \mathbf{B}$ to their discrete counterparts $\overline{\mathbf{A}}, \overline{\mathbf{B}}$. The Mamba architecture adopt the zero-order hold (ZOH) for transformation, which is formulated as follows:
\begin{equation}
\begin{aligned}
    \overline{\mathbf{A}} &= \exp{(\mathbf{\Delta}\mathbf{A})}, \\
    \overline{\mathbf{B}} &= (\mathbf{\Delta} \mathbf{A})^{-1} (\exp{\mathbf{\Delta} \mathbf{A}} - \mathbf{I})\cdot\mathbf{\Delta}\mathbf{B}.
    \end{aligned}
\end{equation}
When the system model has no direct feedthrough, $\mathbf{D}$ is a zero matrix. Thus, we have the discretized version of Equation (\ref{continue_equation}), given as:
\textcolor{black}{\begin{equation}\label{mamba_euqation}
\begin{aligned}
    h_{t} & = \overline{\mathbf{A}}\mathbf{h}_{t-1} + \overline{\mathbf{B}}x_{t}, \\
    z_{t} &= \mathbf{C}\mathbf{h}_{t}.
\end{aligned}
\end{equation}}
We expand the Equation (\ref{mamba_euqation}) and have:
\textcolor{black}{
\begin{equation}
\begin{aligned}
& z_{0}=\mathbf{C} \overline{\mathbf{A}}^0 \overline{\mathbf{B}} u_{0} \\
& z_{1}=\mathbf{C} \overline{\mathbf{A}}^1 \overline{\mathbf{B}} u_{0}+\mathbf{C} \overline{\mathbf{A}}^0 \overline{\mathbf{B}} u_{1} \\
& z_{2}=\mathbf{C} \overline{\mathbf{A}}^2 \overline{\mathbf{B}} u_{0}+\mathbf{C} \overline{\mathbf{A}}^1 \overline{\mathbf{B}} u_{1}+\mathbf{C} \overline{\mathbf{A}}^0 \overline{\mathbf{B}} u_{2}.
\end{aligned}
\end{equation}}
Furthermore, we can rewrite the Equation (\ref{mamba_euqation}) with a global convolution formulation as follows:
\textcolor{black}{\begin{equation}
\begin{aligned}
\overline{\mathbf{K}} & =\left(\mathbf{C} \overline{\mathbf{B}}, \mathbf{C} \overline{\mathbf{A B}}, \ldots, \mathbf{C} \overline{\mathbf{A}}^{L-1} \overline{\mathbf{B}}, \ldots\right) \\
\boldsymbol{z} & =\mathbf{u} * \overline{\mathbf{K}},
\end{aligned}
\end{equation}}
where $L$ is the length of the input sequence, $\overline{\mathbf{K}}$ is a structured convolution kernel, and $*$ denotes the convolution operation.}

\textcolor{black}{\textbf{Selectivity.} Since the parameters $\mathbf{A}$, $\mathbf{B}$, and $\mathbf{C}$ are input-independent, the original SSMs performs poorly in contextual learning. Mamba~\cite{gu2023mamba} further introduces the selective scan mechanism to enable the model to dynamically adjust the $\mathbf{\Delta}$, $\mathbf{B}$, and $\mathbf{C}$ (functions of input) with respect to the inputs. It allows the model to learn dynamic representation and filter relevant information out. In addition, Mamba allows for efficient training through parallel scanning and kernel fusion.}


\begin{figure*}[!t]
\centering
\begin{subfigure}[t]{1.1\columnwidth}
\centerline{\includegraphics[width=\columnwidth]{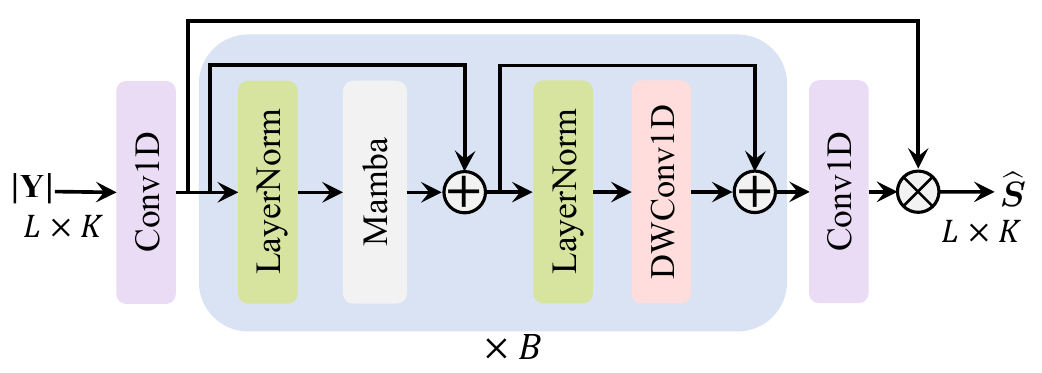}}
\caption{}
\label{fig2:1}
\end{subfigure}
\begin{subfigure}[t]{0.9\columnwidth}
\centerline{\includegraphics[width=1.0\columnwidth]{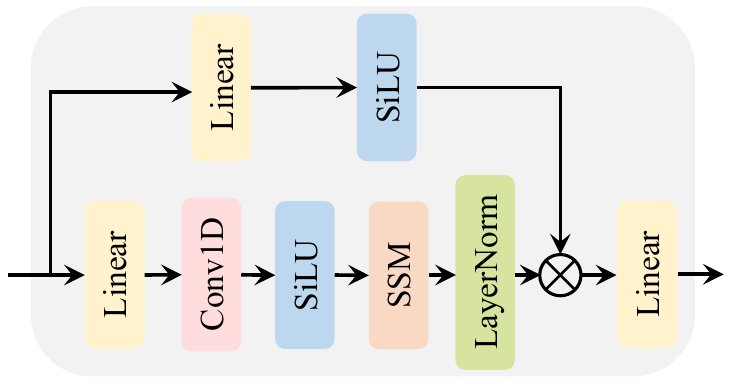}}
\caption{}
\label{fig2:2}
\end{subfigure}
\caption{Diagrams of (a) the overall network architecture of our MambaDC speech enhancement and (b) the Mamba layer. \textcolor{black}{DWConv1D represents the depth-wise 1-D convolution unit. The symbol $\otimes$ represents the element-wise multiplication operation.}}
\label{net_arc}
\end{figure*}

\subsection{Network Architecture}\label{sec:4.1}
\textcolor{black}{Figure~\ref{net_arc}\,(a) illustrates the overall network architecture of our MambaDC speech enhancement model. It takes the STFT magnitude spectrum of a noisy speech as the input, denoted as $|\mathbf{Y}|\in\mathbb{R}^{L\times K}$, with $L$ and $K$ the number of frames and discrete frequency bands, respectively. The input is initially processed by a 1-D convolution layer, which is preactivated by the frame-wise layer normalization (LN)~\cite{tfa} followed by the ReLU activation. This produces a latent representation denoted as $\mathbf{Z}\in\mathbb{R}^{T\times d_{model}}$. Subsequently, $\mathbf{Z}$ passes through stacked $B$ MambaDC layers, with the feature map $\mathbf{H}^{b}\in\mathbb{R}^{T\times d_{model}}$ at the $b$-th layer, $b\in\{1,2,...,B\}$. Each MambaDC layer comprises two sub-layers, i.e., a residual Mamba layer for modeling long-range dependencies and a convolution layer for refining features. Finally, a 1-D convolution layer with sigmoid activation function is used to produce the estimate of T-F mask, $\widehat{\mathbf{M}}$, which is applied to $\mathbf{Y}$ to acquire the estimate the clean spectrum \textcolor{black}{$\widehat{\mathbf{S}}=\mathbf{Y}\cdot \widehat{\mathbf{M}}$}.}

\textcolor{black}{\textbf{MambaDC layer}. As depicted in Figure~\ref{net_arc}\,(a), given the input feature map for the $b$-th layer $\mathbf{H}^{b-1}\in\mathbb{R}^{T\times d_{model}}$, the LN followed by the Mamba layer is used to model the long-range dependencies:
\begin{equation}
    \mathbf{E}^{b} = \text{Mamba}\left(\text{LN}\left(\mathbf{H}^{b-1}\right)\right) + \mathbf{H}^{b-1}.
\end{equation}
Mamba excels at capturing the long-range global interactions, while convolution neural networks capture local interactions effectively. To this end, a 1-D depth-wise convolutional layer is introduced after the residual Mamba layer to extract fine-grained local feature patterns, and the output for the $b$-th layer is given as:
\begin{equation}
    \mathbf{H}^{b} = \text{DWConv}\left(\text{LN}\left(\mathbf{E}^{b}\right)\right) + \mathbf{E}^{b}
\end{equation}
\textcolor{black}{where DWConv refers to the depth-wise convolution layer.} The architecture of Mamba layer is illustrated in Figure~\ref{net_arc}\,(b), the input feature denoted as $\mathbf{F}\in\mathbb{R}^{L\times d_{model}}$ will pass through two parallel branches. The first branch is a linear projection layer followed by the SiLU activation function. The second branch comprises a linear layer, a 1-D depth-wise convolution, SiLU activation function, SSM layer and layer norm in sequence. Afterward, the two branches are combined using the element-wise multiplication. Finally, a linear layer is used to generate the output $\mathbf{F}_{\text{out}}$ with the same shape as the input:}
\textcolor{black}{
\begin{equation}
\begin{aligned}
& \mathbf{M}_{1}  =\text{LN}\left(\text{SSM}\left(\text{SiLU}\left(\text{DWConv}\left(\text{Linear}\left(\mathbf{F}\right)\right)\right)\right)\right), \\
& \mathbf{M}_{2} = \text{SiLU}\left(\text{Linear}\left(\mathbf{F}\right)\right),\\
& \mathbf{M} = \text{Linear}\left(\mathbf{M}_{1}\otimes \mathbf{M}_{2} \right)
\end{aligned}
\end{equation}
where $\otimes$ denotes the element-wise multiplication.}

\section{\textcolor{black}{Experimental Setup}}\label{sec:5}

\subsection{\textcolor{black}{Data}}\label{sec:5.2}
\textcolor{black}{The clean speech and noise data used in our experiments are detailed in this section. For clean speech data in the training set, we use the~\textit{train-clean-100} partition of the Librispeech corpus~\cite{Librispeech}, which contains $28\,539$ utterances (approximately 100 hours) from $125$ female and $126$ male speakers. We collect the training noise recordings from multiple data sources, i.e., the Nonspeech dataset~\cite{Nonspeech}, the RSG-10 dataset~\cite{RSG}, Environmental Noise dataset~\cite{ENV1,ENV2}, the coloured noise signals~\cite{DeepMMSE} (with an $\alpha$ value ranging from $-2$ to $2$ in steps of 0.25), the Urban Sound dataset~\cite{Urban}, the QUT-NOISE dataset~\cite{QUT}, and the noise partition of the MUSAN corpus~\cite{MUSAN}. For evaluation experiments, we exclude two non-stationary (\textit{F16} and \textit{factory welding} from RSG-10 dataset) and two coloured (\textit{street music} from Urban Sound dataset~\cite{Urban} and \textit{voice babble} from the RSG-10 noise dataset~\cite{RSG}) real-world noise recordings from the noise data. The noise recordings in the training set over 30 seconds in duration are split into segments of 30 seconds or less. This creates a noise set comprising a total of $6\,809$ noise recordings, each no longer than 30 seconds.}

\textcolor{black}{For the validation set, we randomly draw $1\,000$ clean speech utterances and noise recordings from the aforementioned speech and noise data to create $1\,000$ clean-noisy pairs, where each speech utterance is degraded by a random section in a noise recording in a random \textcolor{black}{signal-to-noise ratio (SNR)} level sampled from the range $[-20, 10]$ dB (in $1$ dB steps). The clean speech data for testing are taken from the \textit{test-clean-100} partition of the Librispeech corpus. Ten clean speech utterances are randomly picked (without replacement) for each of the excluded four real-world noise sources and are degraded by a random section of the noise recordings at five SNR levels $\mathbf{Q}\!\in\{-5, 0, 5, 10, 15\}$ dB. This produces 200 noisy mixtures for evaluation.}

\textcolor{black}{In addition, we further evaluate the our MambaDC model in VoiceBank+DEMAND dataset (VB-DMD)~\cite{demand}, which is a classical benchmark for monaural speech enhancement. The speech data is taken from VCTK corpus~\cite{6709856}, where $11\,572$ utterances (from $28$ speakers) and $827$ utterances (from $2$ unseen speakers) are used for training and testing, respectively. For the training data, the utterances are degraded by $10$ noise types ($8$ real-word recorded noise types from the DEMAND database~\cite{thiemann2013diverse} and $2$ artificial noise types) at SNRs of $0$, $5$, $10$, and $15$ dB. The SNR levels for testing are $2.5$, $7.5$, $12.$5, and $17.5$ dB. \textcolor{black}{All audio recordings are sampled at a rate of 16 kHz.}}

\subsection{\textcolor{black}{Feature Extraction}}\label{sec:5.1}
\textcolor{black}{The noisy speech utterance is segmented into a set of time frames using a square-root-Hann analysis window of length 512 samples (32 ms) with a hop length of 256 samples (16 ms). A 512-point STFT is applied for each time frame, leading to a 257-point STFT magnitude spectrum as the input to models.}

\subsection{\textcolor{black}{Model Configurations}}
\textcolor{black}{\textcolor{black}{To demonstrate the superiority of the MambaDC model, we employ the original Mamba model~\cite{gu2023mamba,mambaspeech} as a base, with the default hyper-parameters: the state dimension $16$, convolution dimension $4$, and the expansion factor $2$.} Our comparison experiments involve model configurations with $B\!=\!4$ and $B\!=\!7$ Mamba layers, across two commonly used T-F masks, i.e., IRM and PSM. 
\textcolor{black}{The Mamba models with $4$ and $7$ layers are referred to Mamba-4 and Mamba-7. The counterpart MambaDC models are referred to as MambaDC-4, and MambaDC-7.} In addition, we also compare the MambaDC with state-of-the-art (SoTA) backbone networks, i.e., Transformer~\cite{mhanet,zhang2024empirical} and Conformer~\cite{conformer-asr,tfa}.
}

\textcolor{black}{The Transformer backbone includes $4$ stacked Transformer layers with the same parameter configurations, i.e., the number of attention heads \textcolor{black}{$H=8$}, the model dimension $d_{model}=256$, and the inner layer dimension of the feed-forward network (FFN) $d_{f\!f}=1024$. The Conformer backbone consists of $4$ stacked Conformer layers with the parameter configurations as in~\cite{tfaj,conformer-asr}, i.e., the number of attention heads \textcolor{black}{$H=8$}, the model dimension $d_{model}=256$, the inner layer dimension of the feed-forward network (FFN) $d_{f\!f}=1024$, and the convolution kernel size $k=32$. All models are implemented using PyTorch $1.13.0$ and the experiments are run with NVIDIA Tesla V100-32GB graphics processing units (GPUs).}

\subsection{\textcolor{black}{Training Methodology}}

\textcolor{black}{This section details the methodology of training models. We generate noisy mixtures by dynamically mixing clean speech and noise at training time. The clean speech is mixed with a random clip of a randomly picked noise recordings at an SNR level randomly sampled from $-10$ to $20$ dB (in $1$ dB steps). Each mini-batch consists of $10$ speech utterances for one gradient update. The utterances in a mini-batch are zero-padded to match the length of the longest one. We randomly shuffle the order of the speech utterances at the start of each epoch. \textcolor{black}{The mask approximation mean-square error (MSE) is used as the objective function to learn T-F masks.} The~\textit{Adam} optimizer~\cite{Adam} is used to update gradient for all the models, with default hyper-parameters, $\beta_{1}\!=\!0.9$, $\beta_{2}\!=\!0.999$, and a learning rate of $1e^{-3}$. The gradient clipping technology is used to clip the gradients to between $-1$ and $1$~\cite{tfa}. Training is conducted over a total of $150$ epochs. For a fair comparison, the training of all the models employ the same warm-up strategy to adjust the learning rate~\cite{mhanet}:
\begin{equation}
\vspace{-0.2em}
lr = d_{model}^{-0.5}\cdot \textrm{min} \left(num\_step^{-0.5}, num\_step \cdot war\_step^{-1.5}\right)
\end{equation}
where $num\_step$ and $warm\_step$ respectively represent the number of training steps and warm-up training steps. We set $war\_step\!=\!40\,000$ to conduct a warm-up training as in~\cite{mhanet}. \textcolor{black}{The best model is selected using cross-validation.}}

\begin{figure}[!b]
\centering
\begin{subfigure}[t]{0.49\columnwidth}
\centerline{\includegraphics[width=\columnwidth]{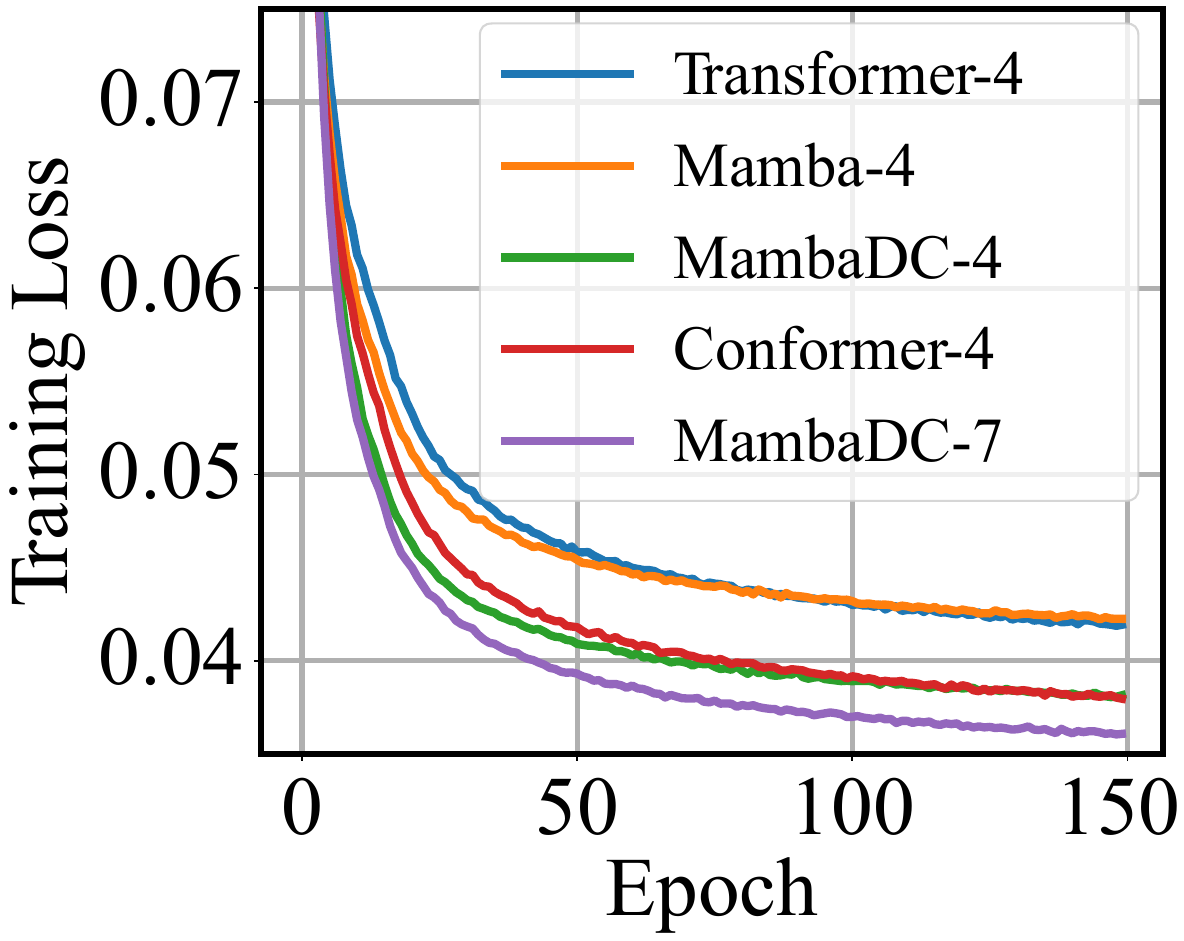}}
\caption{}
\label{fig2:1}
\end{subfigure}
\begin{subfigure}[t]{0.49\columnwidth}
\centerline{\includegraphics[width=1.0\columnwidth]{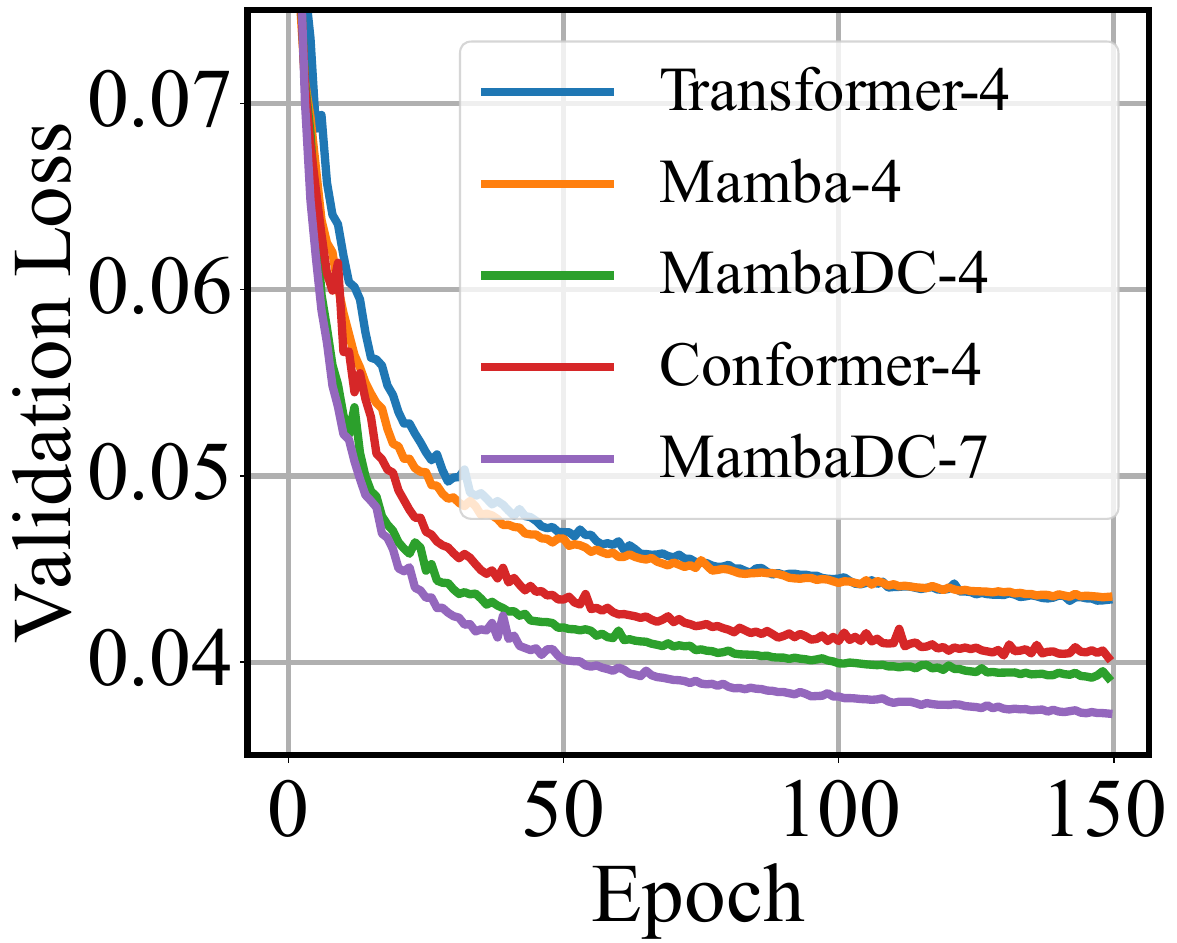}}
\caption{}
\label{fig2:2}
\end{subfigure}
\caption{\textcolor{black}{The curves of the training loss (a) and validation loss (b) of the models on IRM training objective.}}
\label{loss_irm}
\vspace{-1.0em}
\end{figure}

\begin{figure}[!hbtp]
\centering
\begin{subfigure}[t]{0.495\columnwidth}
\centerline{\includegraphics[width=\columnwidth]{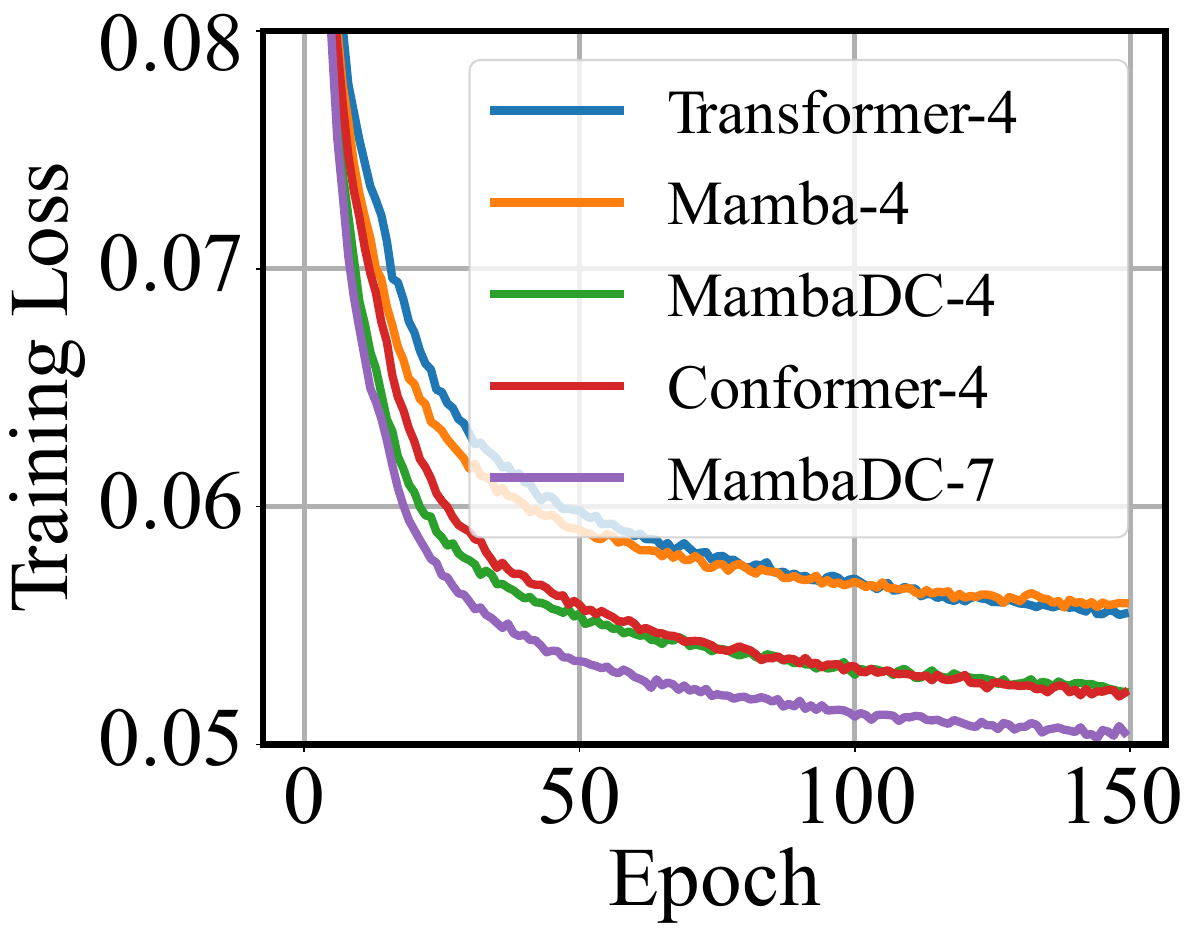}}
\caption{}
\label{fig2:1}
\end{subfigure}
\begin{subfigure}[t]{0.49\columnwidth}
\centerline{\includegraphics[width=1.0\columnwidth]{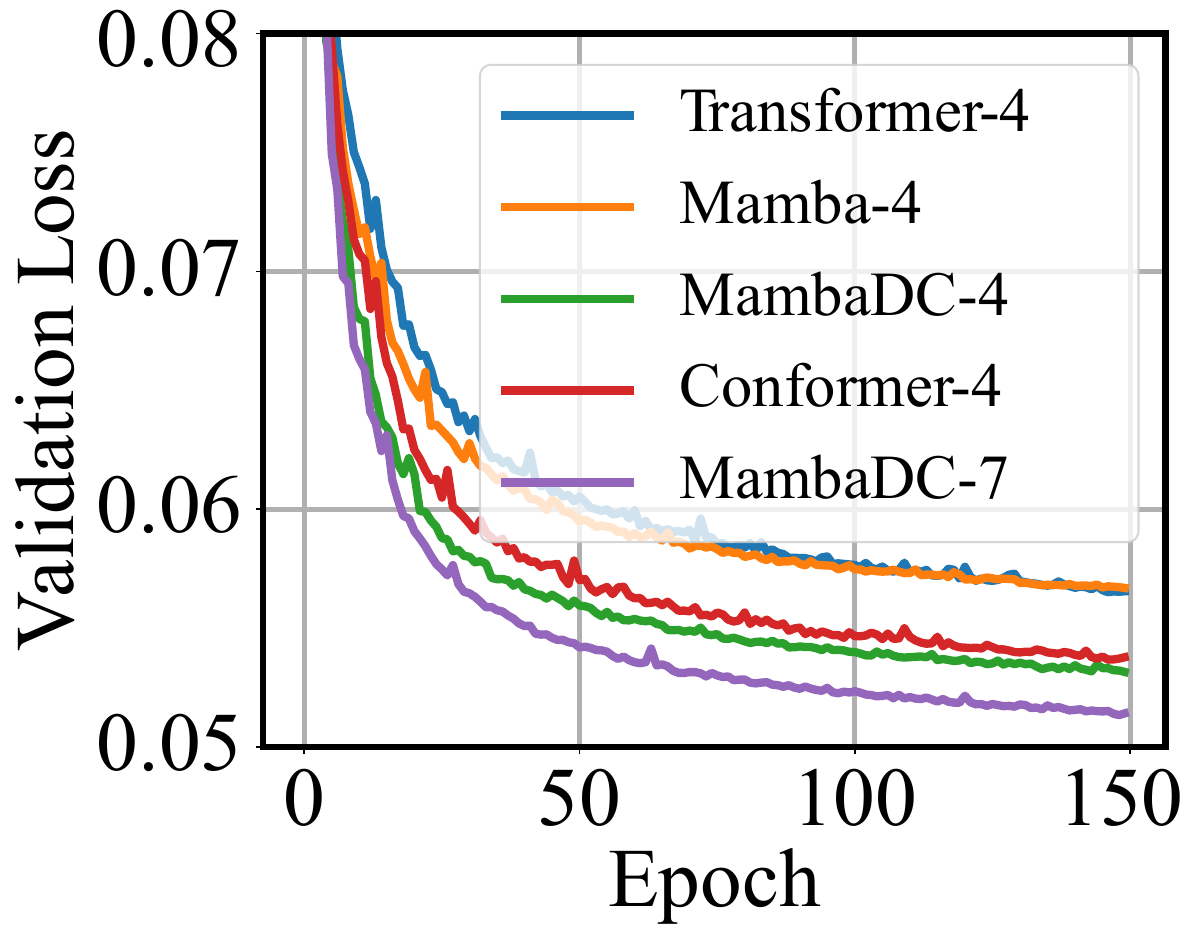}}
\caption{}
\label{loss_psm}
\end{subfigure}
\caption{\textcolor{black}{The curves of the training loss (a) and validation loss (b) of the models on PSM training objective.}}
\label{loss_psm}
\vspace{-1.0em}
\end{figure}

\subsection{\textcolor{black}{Assessment Metrics}}

\textcolor{black}
{Five most commonly used assessment metrics are adopted to evaluate enhanced speech signals, i.e., the wide-band perceptual evaluation of speech quality (WB-PESQ)~\cite{PESQ}, the extended short-time objective intelligibility (ESTOI)~\cite{estoi}, \textcolor{black}{and three composite metrics~\cite{composite}. \textcolor{black}{It should be noted that WB-PESQ typically produces a lower score than the narrow-band counterpart~\cite{restcnsa}.} For all the five metrics, a higher score means better enhancement performance.}
\begin{itemize}
    \item The WB-PESQ assesses the mean opinion score (MOS) for objective speech perceptual quality, with a range from $-0.5$ to $4.5$.
    \item The ESTOI evaluates the MOS of speech intelligibility, typically in the range from $0$ to $1$.
    \textcolor{black}{\item The CSIG composite metric predicts the MOS for the signal distortion (CSIG)~\cite{composite}, with a range from $0$ to $5$.
    \item The CBAK predicts the MOS for the background-noise intrusiveness (CBAK)~\cite{composite}, with a range from $0$ to $5$.
    \item The COVL predicts the MOS for the overall signal quality (COVL)~\cite{composite}, with a range from $0$ to $5$.}
\end{itemize}
}

\begin{table}[!b]
    \centering
    \footnotesize
     \def\arraystretch{1.3}
    \setlength{\tabcolsep}{4.3pt}
    \caption{\textcolor{black}{The evaluation results of the models in terms of WB-PESQ, with IRM as training target. The highest PESQ scores for each test noisy condition are in boldface.}}
    \begin{tabular}{@{}clclllll@{}}
        \toprule[1.25pt]
        &  &  & \multicolumn{5}{c}{\textbf{Input SNR (dB)}} \\  
        \cmidrule{4-8}
        \bf{Noise} & \bf{Network} & \bf{\# Params} & \bf -5 & \bf 0 & \bf 5 & \bf 10 & \bf 15 \\
        \midrule
        \midrule
        
\multirow{9}{*}{\rotatebox[origin=c]{90}{Voice babble}}
& Noisy & \multicolumn{1}{c}{--}  
& 1.07 & 1.12 & 1.23 & 1.47 & 1.89  \\
\cdashline{2-8}




& Transformer-4 \cite{mhanet} & 3.29M
& \textcolor{black}{1.15} & \textcolor{black}{1.36} & \textcolor{black}{1.73} & \textcolor{black}{2.14} & \textcolor{black}{2.54} \\

& Conformer-4~\cite{conformer-asr} & 6.22M 
& \textcolor{black}{1.19} & \textcolor{black}{1.44} & \textcolor{black}{1.86} & \textcolor{black}{2.25} & \textcolor{black}{2.62} \\



& Mamba-4 \cite{mambaspeech} & 1.88M 
& \textcolor{black}{1.17} & \textcolor{black}{1.36} & \textcolor{black}{1.74} & \textcolor{black}{2.15} & \textcolor{black}{2.55} \\

& MambaDC-4 & 1.92M 
& \textcolor{black}{1.19} & \textcolor{black}{1.47} & \textcolor{black}{1.94} & \textcolor{black}{2.33} & \textcolor{black}{2.68} \\

& Mamba-7 \cite{mambaspeech} & 3.20M
& \textcolor{black}{1.19} & \textcolor{black}{1.42} & \textcolor{black}{1.83} & \textcolor{black}{2.23} & \textcolor{black}{2.59} \\

& MambaDC-7 & 3.26M 
& \textcolor{black}{1.23} & \textcolor{black}{1.52} & \textcolor{black}{1.98} & \textcolor{black}{2.34} & \textcolor{black}{2.70} \\

& MambaDC-13 & 5.94M 
& \textcolor{black}{\textbf{1.25}} & \textcolor{black}{\textbf{1.58} }& \textcolor{black}{\textbf{2.04}} & \textcolor{black}{\textbf{2.42}} & \textcolor{black}{\textbf{2.68} }\\

\midrule

\multirow{9}{*}{\rotatebox[origin=c]{90}{Street music}}
& Noisy & \multicolumn{1}{c}{--} 
& 1.03 & 1.05 & 1.10 & 1.25 & 1.56 \\
\cdashline{2-8}




& Transformer-4 \cite{mhanet} & 3.29M
& \textcolor{black}{1.11} & \textcolor{black}{1.27} & \textcolor{black}{1.55} & \textcolor{black}{1.91} & \textcolor{black}{2.25} \\

& Conformer-4~\cite{conformer-asr} & 6.22M 
& \textcolor{black}{1.14} & \textcolor{black}{1.31} & \textcolor{black}{1.62} & \textcolor{black}{2.01} & \textcolor{black}{2.34} \\


& Mamba-4 \cite{mambaspeech} & 1.88M 
& \textcolor{black}{1.10} & \textcolor{black}{1.26} & \textcolor{black}{1.55} & \textcolor{black}{1.91} & \textcolor{black}{2.22} \\

& MambaDC-4 & 1.92M 
& \textcolor{black}{1.14} & \textcolor{black}{1.33} & \textcolor{black}{1.68} & \textcolor{black}{2.08} & \textcolor{black}{2.42} \\

& Mamba-7 \cite{mambaspeech} & 3.20M 
& \textcolor{black}{1.12} & \textcolor{black}{1.30} & \textcolor{black}{1.60} & \textcolor{black}{2.01} & \textcolor{black}{2.35} \\

& MambaDC-7 & 3.26M 
& \textcolor{black}{1.17} & \textcolor{black}{1.39} & \textcolor{black}{1.77} & \textcolor{black}{2.18} & \textcolor{black}{2.51} \\

& MambaDC-13 & 5.94M 
& \textcolor{black}{\textbf{1.20}} & \textcolor{black}{\textbf{1.43}} & \textcolor{black}{\textbf{1.82}} & \textcolor{black}{\textbf{2.22} }& \textcolor{black}{\textbf{2.53}} \\

\midrule

\multirow{9}{*}{\rotatebox[origin=c]{90}{F16}}
& Noisy & \multicolumn{1}{c}{--}  
& 1.04 & 1.06 & 1.11 & 1.27 & 1.58 \\
\cdashline{2-8}




& Transformer-4 \cite{mhanet} & 3.29M
& \textcolor{black}{1.20} & \textcolor{black}{1.44} & \textcolor{black}{1.77} & \textcolor{black}{2.28} & \textcolor{black}{2.60} \\

& Conformer-4~\cite{conformer-asr} & 6.22M 
& \textcolor{black}{1.25} & \textcolor{black}{1.49} & \textcolor{black}{1.81} & \textcolor{black}{2.35} & \textcolor{black}{2.65} \\



& Mamba-4 \cite{mambaspeech} & 1.88M 
& \textcolor{black}{1.19} & \textcolor{black}{1.42} & \textcolor{black}{1.75} & \textcolor{black}{2.23} & \textcolor{black}{2.58} \\

& MambaDC-4 & 1.92M 
& \textcolor{black}{1.28} & \textcolor{black}{1.54} & \textcolor{black}{1.88} & \textcolor{black}{2.38} & \textcolor{black}{2.67} \\

& Mamba-7 \cite{mambaspeech} & 3.20M 
& \textcolor{black}{1.23} & \textcolor{black}{1.47} & \textcolor{black}{1.80} & \textcolor{black}{2.34} & \textcolor{black}{2.64} \\

& MambaDC-7 & 3.26M 
& \textcolor{black}{1.32} & \textcolor{black}{1.57} & \textcolor{black}{1.91} & \textcolor{black}{2.46} & \textcolor{black}{2.71} \\

& MambaDC-13 & 5.94M 
& \textcolor{black}{\textbf{1.37}} & \textcolor{black}{\textbf{1.62}} & \textcolor{black}{\textbf{1.96}} & \textcolor{black}{\textbf{2.46}} & \textcolor{black}{\textbf{2.71}} \\

\midrule

\multirow{9}{*}{\rotatebox[origin=c]{90}{Factory}}
& Noisy & \multicolumn{1}{c}{--}  
& 1.05 & 1.05 & 1.10 & 1.24 & 1.52 \\
\cdashline{2-8}




& Transformer-4 \cite{mhanet} & 3.28M
& \textcolor{black}{1.11} & \textcolor{black}{1.27} & \textcolor{black}{1.54} & \textcolor{black}{1.97} & \textcolor{black}{2.34} \\

& Conformer-4~\cite{conformer-asr} & 6.22M 
& \textcolor{black}{1.15} & \textcolor{black}{1.36} & \textcolor{black}{1.67} & \textcolor{black}{2.12} & \textcolor{black}{2.45} \\



& Mamba-4 \cite{mambaspeech} & 1.88M 
& \textcolor{black}{1.12} & \textcolor{black}{1.29} & \textcolor{black}{1.54} & \textcolor{black}{1.99} & \textcolor{black}{2.36} \\

& MambaDC-4 & 1.92M 
& \textcolor{black}{1.18} & \textcolor{black}{1.41} & \textcolor{black}{1.73} & \textcolor{black}{2.27} & \textcolor{black}{2.44} \\

& Mamba-7 \cite{mambaspeech} & 3.20M 
& \textcolor{black}{1.13} & \textcolor{black}{1.32} & \textcolor{black}{1.59} & \textcolor{black}{2.12} & \textcolor{black}{2.42} \\

& MambaDC-7 & 3.26M 
& \textcolor{black}{1.21} & \textcolor{black}{1.49} & \textcolor{black}{1.85} & \textcolor{black}{2.35} & \textcolor{black}{2.55} \\

& MambaDC-13 & 5.94M 
& \textcolor{black}{\textbf{1.23}} & \textcolor{black}{\textbf{1.54}} & \textcolor{black}{\textbf{1.87} }& \textcolor{black}{\textbf{2.37}} & \textcolor{black}{\textbf{2.58}} \\

\toprule[1.25pt]
    \end{tabular}
    \label{tab1}
\end{table}

\begin{table}[!b]
    \centering
    \footnotesize
     \def\arraystretch{1.3}
    \setlength{\tabcolsep}{4.3pt}
    \caption{\textcolor{black}{The evaluation results of the models in terms of WB-PESQ, with PSM as training target. The highest PESQ scores for each test SNR condition are in boldface.}}
    \begin{tabular}{@{}clclllll@{}}
        \toprule[1.25pt]
        &  &  & \multicolumn{5}{c}{\textbf{Input SNR (dB)}} \\  
        \cmidrule{4-8}
        \bf{Noise} & \bf{Network} & \bf{\# Params} & \bf -5 & \bf 0 & \bf 5 & \bf 10 & \bf 15 \\
        \midrule
        \midrule
        
\multirow{9}{*}{\rotatebox[origin=c]{90}{Voice babble}}
& Noisy & \multicolumn{1}{c}{--}  
& 1.07 & 1.12 & 1.23 & 1.47 & 1.89  \\
\cdashline{2-8}




& Transformer \cite{mhanet} & 3.29M
& \textcolor{black}{1.19} & \textcolor{black}{1.41} & \textcolor{black}{1.79} & \textcolor{black}{2.21} & \textcolor{black}{2.59} \\

& Conformer~\cite{conformer-asr} & 6.22M 
& \textcolor{black}{1.21} & \textcolor{black}{1.48} & \textcolor{black}{1.93} & \textcolor{black}{2.34} & \textcolor{black}{2.67} \\



& Mamba-4 \cite{mambaspeech} & 1.88M 
& \textcolor{black}{1.18} & \textcolor{black}{1.42} & \textcolor{black}{1.82} & \textcolor{black}{2.25} & \textcolor{black}{2.61} \\

& MambaDC-4 & 1.92M 
& \textcolor{black}{1.24} & \textcolor{black}{1.53} & \textcolor{black}{1.98} & \textcolor{black}{2.41} & \textcolor{black}{2.75} \\

& Mamba-7 \cite{mambaspeech} & 3.20M
& \textcolor{black}{1.19} & \textcolor{black}{1.45} & \textcolor{black}{1.91} & \textcolor{black}{2.31} & \textcolor{black}{2.65} \\

& MambaDC-7 & 3.26M 
& \textcolor{black}{1.26} & \textcolor{black}{1.57} & \textcolor{black}{2.04} & \textcolor{black}{2.47} & \textcolor{black}{2.79} \\

& MambaDC-13 & 5.94M 
& \textcolor{black}{\textbf{1.28}} & \textcolor{black}{\textbf{1.63} }& \textcolor{black}{\textbf{2.13}} & \textcolor{black}{\textbf{2.50}} & \textcolor{black}{\textbf{2.86} }\\

\midrule

\multirow{9}{*}{\rotatebox[origin=c]{90}{Street music}}
& Noisy & \multicolumn{1}{c}{--} 
& 1.03 & 1.05 & 1.10 & 1.25 & 1.56 \\
\cdashline{2-8}




& Transformer \cite{mhanet} & 3.29M
& \textcolor{black}{1.13} & \textcolor{black}{1.32} & \textcolor{black}{1.63} & \textcolor{black}{2.03} & \textcolor{black}{2.37} \\

& Conformer~\cite{conformer-asr} & 6.22M 
& \textcolor{black}{1.17} & \textcolor{black}{1.39} & \textcolor{black}{1.76} & \textcolor{black}{2.15} & \textcolor{black}{2.46} \\


& Mamba-4 \cite{mambaspeech} & 1.88M 
& \textcolor{black}{1.13} & \textcolor{black}{1.33} & \textcolor{black}{1.66} & \textcolor{black}{2.08} & \textcolor{black}{2.43} \\

& MambaDC-4 & 1.92M 
& \textcolor{black}{1.17} & \textcolor{black}{1.40} & \textcolor{black}{1.78} & \textcolor{black}{2.23} & \textcolor{black}{2.58} \\

& Mamba-7 \cite{mambaspeech} & 3.20M 
& \textcolor{black}{1.15} & \textcolor{black}{1.35} & \textcolor{black}{1.68} & \textcolor{black}{2.08} & \textcolor{black}{2.41} \\

& MambaDC-7 & 3.26M 
& {1.19} & {1.47} & {1.88} & {2.29} & {2.64} \\

& MambaDC-13 & 5.94M 
& \textcolor{black}{\textbf{1.24}} & \textcolor{black}{\textbf{1.52}} & \textcolor{black}{\textbf{1.96}} & \textcolor{black}{\textbf{2.42} }& \textcolor{black}{\textbf{2.74}} \\

\midrule

\multirow{9}{*}{\rotatebox[origin=c]{90}{F16}}
& Noisy & \multicolumn{1}{c}{--}  
& 1.04 & 1.06 & 1.11 & 1.27 & 1.58 \\
\cdashline{2-8}




& Transformer \cite{mhanet} & 3.29M
& \textcolor{black}{1.28} & \textcolor{black}{1.55} & \textcolor{black}{1.88} & \textcolor{black}{2.46} & \textcolor{black}{2.74} \\

& Conformer~\cite{conformer-asr} & 6.22M 
& \textcolor{black}{1.32} & \textcolor{black}{1.59} & \textcolor{black}{1.93} & \textcolor{black}{2.55} & \textcolor{black}{2.82} \\



& Mamba-4 \cite{mambaspeech} & 1.88M 
& \textcolor{black}{1.27} & \textcolor{black}{1.54} & \textcolor{black}{1.86} & \textcolor{black}{2.44} & \textcolor{black}{2.73} \\

& MambaDC-4 & 1.92M 
& \textcolor{black}{1.34} & \textcolor{black}{1.64} & \textcolor{black}{1.99} & \textcolor{black}{2.59} & \textcolor{black}{2.87} \\

& Mamba-7 \cite{mambaspeech} & 3.20M 
& \textcolor{black}{1.29} & \textcolor{black}{1.57} & \textcolor{black}{1.90} & \textcolor{black}{2.50} & \textcolor{black}{2.75} \\

& MambaDC-7 & 3.26M 
& \textcolor{black}{1.39} & \textcolor{black}{1.70} & \textcolor{black}{2.07} & \textcolor{black}{2.65} & \textcolor{black}{2.91} \\

& MambaDC-13 & 5.94M 
& \textcolor{black}{\textbf{1.41}} & \textcolor{black}{\textbf{1.72}} & \textcolor{black}{\textbf{2.10}} & \textcolor{black}{\textbf{2.71}} & \textcolor{black}{\textbf{2.93}} \\

\midrule

\multirow{9}{*}{\rotatebox[origin=c]{90}{Factory}}
& Noisy & \multicolumn{1}{c}{--}  
& 1.05 & 1.05 & 1.10 & 1.24 & 1.52 \\
\cdashline{2-8}




& Transformer \cite{mhanet} & 3.28M
& \textcolor{black}{1.13} & \textcolor{black}{1.33} & \textcolor{black}{1.66} & \textcolor{black}{2.18} & \textcolor{black}{2.47} \\

& Conformer~\cite{conformer-asr} & 6.22M 
& \textcolor{black}{1.17} & \textcolor{black}{1.40} & \textcolor{black}{1.79} & \textcolor{black}{2.36} & \textcolor{black}{2.56} \\



& Mamba-4 \cite{mambaspeech} & 1.88M 
& \textcolor{black}{1.15} & \textcolor{black}{1.34} & \textcolor{black}{1.68} & \textcolor{black}{2.24} & \textcolor{black}{2.45} \\

& MambaDC-4 & 1.92M 
& \textcolor{black}{1.24} & \textcolor{black}{1.55} & \textcolor{black}{1.90} & \textcolor{black}{2.44} & \textcolor{black}{2.64} \\

& Mamba-7 \cite{mambaspeech} & 3.20M 
& \textcolor{black}{1.19} & \textcolor{black}{1.38} & \textcolor{black}{1.75} & \textcolor{black}{2.28} & \textcolor{black}{2.47} \\

& MambaDC-7 & 3.26M 
& \textcolor{black}{1.29} & \textcolor{black}{1.62} & \textcolor{black}{2.00} & \textcolor{black}{2.54} & \textcolor{black}{2.72} \\

& MambaDC-13 & 5.94M 
& \textcolor{black}{\textbf{1.31}} & \textcolor{black}{\textbf{1.64}} & \textcolor{black}{\textbf{2.04} }& \textcolor{black}{\textbf{2.56}} & \textcolor{black}{\textbf{2.74}} \\

\toprule[1.25pt]
    \end{tabular}
    \label{tab2}
\end{table}

\section{\textcolor{black}{Experiment Results and Analysis}}\label{sec:6}

\subsection{\textcolor{black}{Training and Validation Loss}} 
\textcolor{black}{In Figures~\ref{loss_irm} and \ref{loss_psm}, we give the training and validation loss curves produced by different models on IRM and PSM, respectively, over $150$ training epochs. It can be observed that the loss curves on IRM and PSM showcase a similar trend. Mamba-4 (1.88M) produces similar training and validation loss values compared to Transformer-4 (3.29M). MambaDC-4 (1.92M) and MambaDC-7 (3.26M) demonstrate substantially lower loss values than Transformer-4 and Conformer-4 (6.22M) as well as their Mamba counterparts, respectively, which confirms the efficacy of our network design and the superiority of our MambaDC over prior SoTA architectures. MambaDC-4 also achieves a slightly lower loss value than Conformer-4.}

\begin{table}[!t]
    \centering
    \footnotesize
    \def\arraystretch{1.3}
    \setlength{\tabcolsep}{2.9pt}
    \caption{\textcolor{black}{The evaluation results of the models in terms of ESTOI (in \%), with IRM as training target. The highest ESTOI scores for each SNR condition are in boldface.}}
    \begin{tabular}{@{}clclllll@{}}
        \toprule[1.25pt]
        &  &  & \multicolumn{5}{c}{\textbf{Input SNR (dB)}} \\  
        \cmidrule{4-8}
        Noise & Network & \# Params & -5 & 0 & 5 & 10 & 15 \\
        \midrule
        \midrule
        
\multirow{9}{*}{\rotatebox[origin=c]{90}{Voice babble}}
& Noisy & \multicolumn{1}{c}{--}  
& 28.76 & 44.42 & 60.67 & 74.97 & 85.37  \\
\cdashline{2-8}






& Transformer-4 \cite{mhanet} & 3.29M
& \textcolor{black}{40.59} & \textcolor{black}{59.40} & \textcolor{black}{74.08} & \textcolor{black}{82.96} & \textcolor{black}{87.88} \\

& Conformer-4 \cite{conformer-asr} & 6.22M
& \textcolor{black}{44.42} & \textcolor{black}{63.47} & \textcolor{black}{77.32} & \textcolor{black}{84.45} & \textcolor{black}{88.60} \\

& Mamba-4 \cite{mambaspeech} & 1.88M 
& \textcolor{black}{40.84} & \textcolor{black}{59.67} & \textcolor{black}{74.91} & \textcolor{black}{83.15} & \textcolor{black}{88.03} \\

& MambaDC-4 & 1.92M 
& \textcolor{black}{44.47} & \textcolor{black}{63.64} & \textcolor{black}{77.43} & \textcolor{black}{84.74} & \textcolor{black}{88.75} \\

& Mamba-7 \cite{mambaspeech} & 3.20M
& \textcolor{black}{43.62} & \textcolor{black}{62.20} & \textcolor{black}{76.46} & \textcolor{black}{84.01} & \textcolor{black}{88.35} \\

& MambaDC-7 & 3.26M 
& \textcolor{black}{48.75} & \textcolor{black}{65.57} & \textcolor{black}{78.72} & \textcolor{black}{85.04} & \textcolor{black}{89.08} \\

& MambaDC-13 & 5.94M 
& \textcolor{black}{\textbf{50.04}} & \textcolor{black}{\textbf{67.74}} & \textcolor{black}{\textbf{79.55}} & \textcolor{black}{\textbf{85.61}} & \textcolor{black}{\textbf{89.30}} \\

\midrule

\multirow{9}{*}{\rotatebox[origin=c]{90}{Street music}}
& Noisy & \multicolumn{1}{c}{--} 
& 30.39 & 44.03 & 58.15 & 71.13 & 81.80 \\
\cdashline{2-8}






& Transformer-4 \cite{mhanet} & 3.29M
& \textcolor{black}{42.21} & \textcolor{black}{58.79} & \textcolor{black}{72.24} & \textcolor{black}{81.21} & \textcolor{black}{86.14} \\

& Conformer-4 \cite{conformer-asr} & 6.22M
& \textcolor{black}{45.23} & \textcolor{black}{61.44} & \textcolor{black}{73.54} & \textcolor{black}{82.38} & \textcolor{black}{86.75} \\

& Mamba-4 \cite{mambaspeech} & 1.88M 
& \textcolor{black}{41.61} & \textcolor{black}{58.63} & \textcolor{black}{71.69} & \textcolor{black}{81.08} & \textcolor{black}{85.83} \\

& MambaDC-4 & 1.92M 
& \textcolor{black}{46.35} & \textcolor{black}{61.53} & \textcolor{black}{74.10} & \textcolor{black}{82.48} & \textcolor{black}{86.63} \\

& Mamba-7 \cite{mambaspeech} & 3.20M
& \textcolor{black}{44.55} & \textcolor{black}{60.50} & \textcolor{black}{73.19} & \textcolor{black}{82.04} & \textcolor{black}{86.47} \\

& MambaDC-7 & 3.26M 
& \textcolor{black}{48.61} & \textcolor{black}{63.95} & \textcolor{black}{75.47} & \textcolor{black}{83.43} & \textcolor{black}{87.08} \\

& MambaDC-13 & 5.94M 
& \textcolor{black}{\textbf{50.40}} & \textcolor{black}{\textbf{65.37} }& \textcolor{black}{\textbf{76.50}} & \textcolor{black}{\textbf{84.08}} & \textcolor{black}{\textbf{87.63}} \\

\midrule

\multirow{9}{*}{\rotatebox[origin=c]{90}{F16}}
& Noisy & \multicolumn{1}{c}{--}  
& 27.45 & 41.89 & 56.70 & 70.27 & 81.30 \\
\cdashline{2-8}






& Transformer-4 \cite{mhanet} & 3.29M
& \textcolor{black}{46.41} & \textcolor{black}{61.86} & \textcolor{black}{73.64} & \textcolor{black}{84.27} & \textcolor{black}{87.00} \\

& Conformer-4 \cite{conformer-asr} & 6.22M
& \textcolor{black}{50.00} & \textcolor{black}{64.29} & \textcolor{black}{75.08} & \textcolor{black}{85.59} & \textcolor{black}{87.71} \\

& Mamba-4 \cite{mambaspeech} & 1.88M 
& \textcolor{black}{46.60} & \textcolor{black}{61.62} & \textcolor{black}{73.34} & \textcolor{black}{84.16} & \textcolor{black}{86.82} \\

& MambaDC-4 & 1.92M 
& \textcolor{black}{51.80} & \textcolor{black}{65.65} & \textcolor{black}{75.87} & \textcolor{black}{85.83} & \textcolor{black}{87.78} \\

& Mamba-7 \cite{mambaspeech} & 3.20M
& \textcolor{black}{49.14} & \textcolor{black}{63.52} & \textcolor{black}{74.22} & \textcolor{black}{84.88} & \textcolor{black}{87.14} \\

& MambaDC-7 & 3.26M 
& \textcolor{black}{54.92} & \textcolor{black}{67.71} & \textcolor{black}{76.83} & \textcolor{black}{86.43} & \textcolor{black}{88.07} \\

& MambaDC-13 & 5.94M 
& \textcolor{black}{\textbf{56.34}} & \textcolor{black}{\textbf{68.91}} & \textcolor{black}{\textbf{77.96}} & \textcolor{black}{\textbf{87.01} }& \textcolor{black}{\textbf{88.54} }\\

\midrule

\multirow{9}{*}{\rotatebox[origin=c]{90}{Factory}}
& Noisy & \multicolumn{1}{c}{--}  
& 25.03 & 38.45 & 53.30 & 68.09 & 80.42 \\
\cdashline{2-8}






& Transformer \cite{mhanet} & 3.29M
& \textcolor{black}{38.70} & \textcolor{black}{57.02} & \textcolor{black}{69.85} & \textcolor{black}{82.32} & \textcolor{black}{85.85} \\

& Conformer-4 \cite{conformer-asr} & 6.22M
& \textcolor{black}{42.91} & \textcolor{black}{60.18} & \textcolor{black}{72.13} & \textcolor{black}{83.46} & \textcolor{black}{86.50} \\

& Mamba-4 \cite{mambaspeech} & 1.88M 
& \textcolor{black}{39.03} & \textcolor{black}{57.25} & \textcolor{black}{70.21} & \textcolor{black}{82.51} & \textcolor{black}{86.06} \\

& MambaDC-4 & 1.92M 
& \textcolor{black}{44.12} & \textcolor{black}{62.24} & \textcolor{black}{73.70} & \textcolor{black}{84.61} & \textcolor{black}{87.23} \\

& Mamba-7 \cite{mambaspeech} & 3.20M
& \textcolor{black}{41.05} & \textcolor{black}{59.66} & \textcolor{black}{71.64} & \textcolor{black}{83.60} & \textcolor{black}{86.48} \\

& MambaDC-7 & 3.26M 
& \textcolor{black}{46.88} & \textcolor{black}{64.36} & \textcolor{black}{74.93} & \textcolor{black}{85.32} & \textcolor{black}{87.50} \\

& MambaDC-13 & 5.94M 
& \textcolor{black}{\textbf{49.04}} & \textcolor{black}{\textbf{66.18}} & \textcolor{black}{\textbf{75.64}} & \textcolor{black}{\textbf{85.69}} & \textcolor{black}{\textbf{87.40}} \\

\toprule[1.25pt]
    \end{tabular}
    \label{tab3}
\end{table}

\begin{table}[!t]
    \centering
    \footnotesize
    \def\arraystretch{1.3}
    \setlength{\tabcolsep}{2.9pt}
    \caption{\textcolor{black}{The evaluation results of the models in terms of ESTOI (in \%), with PSM as training target. The highest ESTOI scores for each noisy condition are in boldface.}}
    \begin{tabular}{@{}clclllll@{}}
        \toprule[1.25pt]
        &  &  & \multicolumn{5}{c}{\textbf{Input SNR (dB)}} \\  
        \cmidrule{4-8}
        \bf Noise & \bf Network & \# \bf Params & \bf -5 & \bf 0 & \bf 5 & \bf 10 & \bf 15 \\
        \midrule
        \midrule
        
\multirow{9}{*}{\rotatebox[origin=c]{90}{Voice babble}}
& Noisy & \multicolumn{1}{c}{--}  
& 28.76 & 44.42 & 60.67 & 74.97 & 85.37  \\
\cdashline{2-8}






& Transformer-4 \cite{mhanet} & 3.29M
& \textcolor{black}{40.77} & \textcolor{black}{59.81} & \textcolor{black}{74.86} & \textcolor{black}{83.36} & \textcolor{black}{88.07} \\

& Conformer-4 \cite{conformer-asr} & 6.22M
& \textcolor{black}{44.51} & \textcolor{black}{63.47} & \textcolor{black}{77.14} & \textcolor{black}{84.55} & \textcolor{black}{88.71} \\

& Mamba-4 \cite{mambaspeech} & 1.88M 
& \textcolor{black}{41.26} & \textcolor{black}{60.25} & \textcolor{black}{75.42} & \textcolor{black}{83.54} & \textcolor{black}{88.36} \\

& MambaDC-4 & 1.92M 
& \textcolor{black}{45.13} & \textcolor{black}{63.67} & \textcolor{black}{78.11} & \textcolor{black}{84.94} & \textcolor{black}{88.96} \\

& Mamba-7 \cite{mambaspeech} & 3.20M
& \textcolor{black}{42.27} & \textcolor{black}{61.73} & \textcolor{black}{76.37} & \textcolor{black}{84.05} & \textcolor{black}{88.62} \\

& MambaDC-7 & 3.26M 
& \textcolor{black}{47.83} & \textcolor{black}{65.63} & \textcolor{black}{79.04} & \textcolor{black}{85.47} & \textcolor{black}{89.22} \\

& MambaDC-13 & 5.94M 
& \textcolor{black}{\textbf{50.25}} & \textcolor{black}{\textbf{67.58}} & \textcolor{black}{\textbf{80.01}} & \textcolor{black}{\textbf{85.86}} & \textcolor{black}{\textbf{89.46}} \\

\midrule

\multirow{9}{*}{\rotatebox[origin=c]{90}{Street music}}
& Noisy & \multicolumn{1}{c}{--} 
& 30.39 & 44.03 & 58.15 & 71.13 & 81.80 \\
\cdashline{2-8}






& Transformer-4 \cite{mhanet} & 3.29M
& \textcolor{black}{41.27} & \textcolor{black}{58.49} & \textcolor{black}{72.07} & \textcolor{black}{81.58} & \textcolor{black}{86.22} \\

& Conformer-4 \cite{conformer-asr} & 6.22M
& \textcolor{black}{45.21} & \textcolor{black}{61.39} & \textcolor{black}{73.84} & \textcolor{black}{82.49} & \textcolor{black}{86.76} \\

& Mamba-4 \cite{mambaspeech} & 1.88M 
& \textcolor{black}{42.42} & \textcolor{black}{59.09} & \textcolor{black}{72.53} & \textcolor{black}{81.81} & \textcolor{black}{86.30} \\

& MambaDC-4 & 1.92M 
& \textcolor{black}{45.33} & \textcolor{black}{62.25} & \textcolor{black}{73.96} & \textcolor{black}{82.74} & \textcolor{black}{86.72} \\

& Mamba-7 \cite{mambaspeech} & 3.20M
& \textcolor{black}{43.77} & \textcolor{black}{60.55} & \textcolor{black}{73.19} & \textcolor{black}{82.08} & \textcolor{black}{86.51} \\

& MambaDC-7 & 3.26M 
& \textcolor{black}{46.38} & \textcolor{black}{63.95} & \textcolor{black}{75.19} & \textcolor{black}{83.69} & \textcolor{black}{87.22} \\

& MambaDC-13 & 5.94M 
& \textcolor{black}{\textbf{49.79}} & \textcolor{black}{\textbf{64.77} }& \textcolor{black}{\textbf{76.18}} & \textcolor{black}{\textbf{84.24}} & \textcolor{black}{\textbf{87.77}} \\

\midrule

\multirow{9}{*}{\rotatebox[origin=c]{90}{F16}}
& Noisy & \multicolumn{1}{c}{--}  
& 27.45 & 41.89 & 56.70 & 70.27 & 81.30 \\
\cdashline{2-8}






& Transformer-4 \cite{mhanet} & 3.29M
& \textcolor{black}{46.51} & \textcolor{black}{62.80} & \textcolor{black}{74.23} & \textcolor{black}{84.69} & \textcolor{black}{87.03} \\

& Conformer-4 \cite{conformer-asr} & 6.22M
& \textcolor{black}{50.34} & \textcolor{black}{64.73} & \textcolor{black}{75.46} & \textcolor{black}{85.73} & \textcolor{black}{87.75} \\

& Mamba-4 \cite{mambaspeech} & 1.88M 
& \textcolor{black}{47.53} & \textcolor{black}{62.97} & \textcolor{black}{73.98} & \textcolor{black}{84.52} & \textcolor{black}{86.99} \\

& MambaDC-4 & 1.92M 
& \textcolor{black}{51.16} & \textcolor{black}{65.82} & \textcolor{black}{76.16} & \textcolor{black}{86.21} & \textcolor{black}{88.03} \\

& Mamba-7 \cite{mambaspeech} & 3.20M
& \textcolor{black}{49.22} & \textcolor{black}{63.74} & \textcolor{black}{74.87} & \textcolor{black}{85.12} & \textcolor{black}{87.45} \\

& MambaDC-7 & 3.26M 
& \textcolor{black}{54.63} & \textcolor{black}{68.19} & \textcolor{black}{77.53} & \textcolor{black}{86.78} & \textcolor{black}{88.51} \\

& MambaDC-13 & 5.94M 
& \textcolor{black}{\textbf{56.20}} & \textcolor{black}{\textbf{69.34}} & \textcolor{black}{\textbf{78.34}} & \textcolor{black}{\textbf{87.50} }& \textcolor{black}{\textbf{88.65} }\\

\midrule

\multirow{9}{*}{\rotatebox[origin=c]{90}{Factory}}
& Noisy & \multicolumn{1}{c}{--}  
& 25.03 & 38.45 & 53.30 & 68.09 & 80.42 \\
\cdashline{2-8}






& Transformer \cite{mhanet} & 3.29M
& \textcolor{black}{37.91} & \textcolor{black}{56.73} & \textcolor{black}{70.43} & \textcolor{black}{82.70} & \textcolor{black}{85.93} \\

& Conformer-4 \cite{conformer-asr} & 6.22M
& \textcolor{black}{41.75} & \textcolor{black}{60.72} & \textcolor{black}{72.46} & \textcolor{black}{84.13} & \textcolor{black}{86.89} \\

& Mamba-4 \cite{mambaspeech} & 1.88M 
& \textcolor{black}{38.84} & \textcolor{black}{57.30} & \textcolor{black}{70.84} & \textcolor{black}{83.27} & \textcolor{black}{86.42} \\

& MambaDC-4 & 1.92M 
& \textcolor{black}{43.61} & \textcolor{black}{62.20} & \textcolor{black}{73.68} & \textcolor{black}{84.77} & \textcolor{black}{87.45} \\

& Mamba-7 \cite{mambaspeech} & 3.20M
& \textcolor{black}{40.59} & \textcolor{black}{59.87} & \textcolor{black}{72.01} & \textcolor{black}{84.03} & \textcolor{black}{86.72} \\

& MambaDC-7 & 3.26M 
& \textcolor{black}{46.76} & \textcolor{black}{64.99} & \textcolor{black}{75.61} & \textcolor{black}{85.81} & \textcolor{black}{87.67} \\

& MambaDC-13 & 5.94M 
& \textcolor{black}{\textbf{48.44}} & \textcolor{black}{\textbf{65.58}} & \textcolor{black}{\textbf{75.85}} & \textcolor{black}{\textbf{85.97}} & \textcolor{black}{\textbf{88.09}} \\

\toprule[1.25pt]
    \end{tabular}
    \vspace{-0.5em}
    \label{tab4}
\end{table}

\subsection{\textcolor{black}{Experiment on Enhanceent Performance}}

\textcolor{black}{Table~\ref{tab1} and Table~\ref{tab2} respectively showcase the comparison results among different models in terms of WB-PESQ scores with IRM and PSM as the training objectives, across five SNR conditions. The highest WB-PESQ scores for each SNR and noise condition are denoted in bold. We can observe that all the models improve the WB-PESQ scores over unprocessed noisy mixtures by a large margin. In the case of $10$ dB SNR (on PSM), Transformer, Conformer, Mamba-4, and MambaDC-4 provide $0.74$, $0.87$, $0.78$, and $0.94$ WB-PESQ gains over noisy input, respectively. Among IRM and PSM, the latter exhibits better WB-PESQ results across different models. In the $5$ dB SNR case (\textit{street music} noise), for instance, our MambaDC-4 with PSM improves WB-PESQ over MambaDC-4 with IRM by 0.17.}


\textcolor{black}{In comparison to Mamba, the MambaDC consistently exhibits significant WB-PESQ gains with fewer parameter numbers across all SNR and noise conditions, confirming the superiority of the MambaDC architecture. In the case of the \textit{F16} noise at $5$ dB SNR, the MambaDC-4 (\textcolor{black}{1.92M}) outperforms Mamba-7 (\textcolor{black}{3.20M}) by 0.14 and 0.15 WB-PESQ scores on IRM and PSM, respectively. Moreover, we observe that MambaDC exhibits obvious superiority in WB-PESQ compared to SoTA Transformer and Conformer networks. Taking the case of the \textit{Voice babble} noise at $10$ dB SNR, MambaDC-7 (\textcolor{black}{3.20M}) and MambaDC-13 (\textcolor{black}{5.94M}) respectively provide 0.2 and 0.17, and 0.26 and 0.16 PESQ gains over Transformer (\textcolor{black}{3.29M}) and Conformer (\textcolor{black}{6.22M}) on IRM and PSM. MambaDC-4 (\textcolor{black}{1.88M}) also exhibits higher WB-PESQ scores than Transformer and Conformer.}


\begin{table}[!t]
    \centering
    \footnotesize
    \def\arraystretch{1.1}
    \setlength{\tabcolsep}{5.5pt}
    \caption{\textcolor{black}{The average CSIG scores (across all the four noise conditions) of the models for each SNR level and the highest CSIG scores are in boldface.}}
    \label{tab:csig}
    \begin{tabular}{@{}clccccc@{}}
        \toprule[1pt]
        &  & \multicolumn{5}{c}{\textbf{Input SNR (dB)}}\\  
        \cline{3-7}
        \bf Target & \bf Network & \bf -5 & \bf 0 & \bf 5 & \bf 10 & \bf 15 \\
        \hline
        -- & Noisy & 1.49 & 1.80 & 2.20 & 2.65 & 3.16 \\
        \hline
        \hline
        \multirow{7}{*}{IRM}
        & Transformer~\cite{mhanet} & \textcolor{black}{2.15} & \textcolor{black}{2.62} & \textcolor{black}{3.06} & \textcolor{black}{3.53} & \textcolor{black}{3.82}  \\
        & Conformer~\cite{conformer-asr}  & \textcolor{black}{2.26} & \textcolor{black}{2.75} & \textcolor{black}{3.19} & \textcolor{black}{3.67} & \textcolor{black}{3.91}  \\
        & Mamba-4~\cite{mambaspeech} & \textcolor{black}{2.15} & \textcolor{black}{2.61} & \textcolor{black}{3.06} & \textcolor{black}{3.53} & \textcolor{black}{3.82}  \\ 
        & MambaDC-4 & \textcolor{black}{2.30} & \textcolor{black}{2.81} & \textcolor{black}{3.25} & \textcolor{black}{3.72} & \textcolor{black}{3.94}  \\
        & Mamba-7~\cite{mambaspeech} & \textcolor{black}{2.23} & \textcolor{black}{2.72} & \textcolor{black}{3.16} & \textcolor{black}{3.63} & \textcolor{black}{3.91}  \\
        & MambaDC-7 & \textcolor{black}{2.39} & \textcolor{black}{2.89} & \textcolor{black}{3.35} & \textcolor{black}{3.79} & \textcolor{black}{4.01}  \\
        & MambaDC-13 & \textcolor{black}{2.46} & \textcolor{black}{2.95} & \textcolor{black}{3.40} & \textcolor{black}{3.83} & \textcolor{black}{4.01}  \\
        \hline
        
        \multirow{7}{*}{PSM}
        & Transformer~\cite{mhanet}         & \textcolor{black}{2.13} & \textcolor{black}{2.62} & \textcolor{black}{3.08} & \textcolor{black}{3.57} & \textcolor{black}{3.84}  \\
        & Conformer~\cite{conformer-asr}    & \textcolor{black}{2.25} & \textcolor{black}{2.74} & \textcolor{black}{3.21} & \textcolor{black}{3.70} & \textcolor{black}{3.92}  \\
        & Mamba-4~\cite{mambaspeech}  & \textcolor{black}{2.15} & \textcolor{black}{2.64} & \textcolor{black}{3.11} & \textcolor{black}{3.62} & \textcolor{black}{3.88}  \\
        & MambaDC-4 & \textcolor{black}{2.26} & \textcolor{black}{2.79} & \textcolor{black}{3.27} & \textcolor{black}{3.75} & \textcolor{black}{3.97}  \\
        & Mamba-7~\cite{mambaspeech} & \textcolor{black}{2.21} & \textcolor{black}{2.69} & \textcolor{black}{3.17} & \textcolor{black}{3.66} & \textcolor{black}{3.88}  \\
        & MambaDC-7       & \textcolor{black}{2.38} & \textcolor{black}{2.91} & \textcolor{black}{3.37} & \textcolor{black}{3.84} & \textcolor{black}{4.04}  \\
        & MambaDC-13      & \textcolor{black}{2.45} & \textcolor{black}{2.96} & \textcolor{black}{3.42} & \textcolor{black}{3.87} & \textcolor{black}{4.06}  \\
        \toprule[1.0pt]
    \end{tabular}
    \label{tab5}
\end{table}
\begin{table}[!tbp]
    \centering
    \scriptsize
    \def\arraystretch{1.1}
    \setlength{\tabcolsep}{5.5pt}
    \footnotesize	
    \caption{\textcolor{black}{The average CBAK scores (across all the four noise conditions) of the models for each SNR level and the best CBAK scores are highlighted in boldface.}}
    \label{tab:cbak}
    \begin{tabular}{@{}clccccc@{}}
        \toprule[1pt]
        &  & \multicolumn{5}{c}{\textbf{Input SNR (dB)}}\\  
        \cline{3-7}
        \bf Target & \bf Network & \bf -5 & \bf 0 & \bf 5 & \bf 10 & \bf 15 \\
        \hline
        -- & Noisy & 1.15 & 1.40 & 1.58 & 2.13 & 2.61 \\
        \hline
        \hline
        
        \multirow{7}{*}{IRM}

         & Transformer~\cite{mhanet}       & \textcolor{black}{1.63} & \textcolor{black}{1.95} & \textcolor{black}{2.27} & \textcolor{black}{2.74} & \textcolor{black}{2.92}  \\
        & Conformer~\cite{conformer-asr}  & \textcolor{black}{1.72} & \textcolor{black}{2.03} & \textcolor{black}{2.36} & \textcolor{black}{2.83} & \textcolor{black}{3.00}  \\
         & Mamba-4~\cite{mambaspeech} & \textcolor{black}{1.63} & \textcolor{black}{1.95} & \textcolor{black}{2.26} & \textcolor{black}{2.73} & \textcolor{black}{2.92}  \\
        & MambaDC-4    & \textcolor{black}{1.72} & \textcolor{black}{2.05} & \textcolor{black}{2.39} & \textcolor{black}{2.85} & \textcolor{black}{3.00}  \\
        & Mamba-7~\cite{mambaspeech} & \textcolor{black}{1.68} & \textcolor{black}{2.00} & \textcolor{black}{2.32} & \textcolor{black}{2.82} & \textcolor{black}{2.98}  \\
        & MambaDC-7    & \textcolor{black}{1.77} & \textcolor{black}{2.11} & \textcolor{black}{2.45} & \textcolor{black}{2.92} & \textcolor{black}{3.06}  \\
        & MambaDC-13   & \textcolor{black}{1.81} & \textcolor{black}{2.15} & \textcolor{black}{2.49} & \textcolor{black}{2.94} & \textcolor{black}{3.05}  \\
        
        \hline

        \multirow{7}{*}{PSM}

        & Transformer~\cite{mhanet}       & \textcolor{black}{1.66} & \textcolor{black}{2.00} & \textcolor{black}{2.33} & \textcolor{black}{2.82} & \textcolor{black}{2.97}  \\
        & Conformer~\cite{conformer-asr}  & \textcolor{black}{1.75} & \textcolor{black}{2.08} & \textcolor{black}{2.43} & \textcolor{black}{2.92} & \textcolor{black}{3.05}  \\
        & Mamba-4~\cite{mambaspeech} & \textcolor{black}{1.66} & \textcolor{black}{2.00} & \textcolor{black}{2.34} & \textcolor{black}{2.85} & \textcolor{black}{3.00}  \\
        & MambaDC-4 & \textcolor{black}{1.74} & \textcolor{black}{2.11} & \textcolor{black}{2.46} & \textcolor{black}{2.96} & \textcolor{black}{3.33}  \\
        & Mamba-7~\cite{mambaspeech} & \textcolor{black}{1.73} & \textcolor{black}{2.05} & \textcolor{black}{2.40} & \textcolor{black}{2.89} & \textcolor{black}{3.20}  \\
        & MambaDC-7 & \textcolor{black}{1.80} & \textcolor{black}{2.16} & \textcolor{black}{2.52} & \textcolor{black}{3.01} & \textcolor{black}{3.37}  \\
        & MambaDC-13 & \textcolor{black}{1.85} & \textcolor{black}{2.21} & \textcolor{black}{2.56} & \textcolor{black}{3.05} & \textcolor{black}{3.16}  \\
        \toprule[1pt]
    \end{tabular}
    \label{tab6}
\end{table}

\textcolor{black}{Tables \ref{tab3} and \ref{tab4} respectively present the comparison results among different models in ESTOI (in \%) on IRM and PSM. The highest score for each SNR and noise condition is denoted in boldface. A similar performance trend to that in Tables \ref{tab1} and \ref{tab2} is observed in Tables \ref{tab3} and \ref{tab4}. We observe that all the models substantially improve the ESTOI scores over noisy inputs. Again, the MambaDC demonstrates significant ESTOI improvements to the Mamba base model across IRM and PSM. In the case of \textit{Factory} noise at $0$ dB SNR, for instance, MambaDC-4 outperforms Mamba-7 by $2.58\%$ and $3.02\%$ ESTOI scores on IRM and PSM, respectively. Similarly, obvious performance superiority of MambaDC over Transformer and Conformer networks is observed. Taking the case of \textit{Street music} noise at $0$ dB SNR, MambaDC-7 and MambaDC-13 respectively provide $5.16\%$ and $3.93\%$, and $5.46\%$ and $3.38\%$ ESTOI improvements over Transformer and Conformer on IRM and PSM. Overall, MambaDC-4 (1.88 M) shows better ESTOI scores compared to Transformer and Conformer models.}


\begin{table}[!t]
    \centering
    \def\arraystretch{1.1}
    \setlength{\tabcolsep}{5.5pt}
    \footnotesize	
    \caption{\textcolor{black}{The average COVL scores (across all the four noise conditions) of the models for each SNR level and the best COVL scores are highlighted in boldface.}}
    \label{tab:covl}
    \begin{tabular}{@{}clccccc@{}}
        \toprule[1pt]
        &  & \multicolumn{5}{c}{\textbf{Input SNR (dB)}}\\  
        \cline{3-7}
        \bf Target & \bf Network & \bf -5 & \bf 0 & \bf 5 & \bf 10 & \bf 15 \\
        \hline
        -- & Noisy & 1.17 & 1.32 & 1.58 & 1.92 & 2.37 \\
        \hline
        \hline
        \multirow{7}{*}{IRM}

        & Transformer~\cite{mhanet} & \textcolor{black}{1.55} & \textcolor{black}{1.91} & \textcolor{black}{2.31} & \textcolor{black}{2.78} & \textcolor{black}{3.10}  \\
        & Conformer~\cite{conformer-asr}  & \textcolor{black}{1.64} & \textcolor{black}{2.01} & \textcolor{black}{2.43} & \textcolor{black}{2.91} & \textcolor{black}{3.20}  \\
        & Mamba-4~\cite{mambaspeech} & \textcolor{black}{1.55} & \textcolor{black}{1.90} & \textcolor{black}{2.30} & \textcolor{black}{2.77} & \textcolor{black}{3.11}  \\
        & MambaDC-4 & \textcolor{black}{1.65} & \textcolor{black}{2.06} & \textcolor{black}{2.48} & \textcolor{black}{2.96} & \textcolor{black}{3.23}  \\
        & Mamba-7~\cite{mambaspeech} & \textcolor{black}{1.61} & \textcolor{black}{1.99} & \textcolor{black}{2.39} & \textcolor{black}{2.88} & \textcolor{black}{3.19}  \\
        & MambaDC-7 & \textcolor{black}{1.72} & \textcolor{black}{2.14} & \textcolor{black}{2.57} & \textcolor{black}{3.05} & \textcolor{black}{3.31}  \\
        & MambaDC-13 & \textcolor{black}{1.78} & \textcolor{black}{2.19} & \textcolor{black}{2.62} & \textcolor{black}{3.08} & \textcolor{black}{3.31}  \\
        \hline
        
        \multirow{7}{*}{PSM}

        & Transformer~\cite{mhanet} & \textcolor{black}{1.57} & \textcolor{black}{1.94} & \textcolor{black}{2.35} & \textcolor{black}{2.87} & \textcolor{black}{3.17} \\
        & Conformer~\cite{conformer-asr}  & \textcolor{black}{1.64} & \textcolor{black}{2.04} & \textcolor{black}{2.49} & \textcolor{black}{3.00} & \textcolor{black}{3.25} \\
        & Mamba-4~\cite{mambaspeech} & \textcolor{black}{1.56} & \textcolor{black}{1.95} & \textcolor{black}{2.38} & \textcolor{black}{2.91} & \textcolor{black}{3.19} \\
        & MambaDC-4 & \textcolor{black}{1.64} & \textcolor{black}{2.09} & \textcolor{black}{2.54} & \textcolor{black}{3.06} & \textcolor{black}{3.33}  \\
        & Mamba-7~\cite{mambaspeech} & \textcolor{black}{1.61} & \textcolor{black}{1.99} & \textcolor{black}{2.44} & \textcolor{black}{2.95} & \textcolor{black}{3.20} \\
        & MambaDC-7 & \textcolor{black}{1.73} & \textcolor{black}{2.18} & \textcolor{black}{2.64} & \textcolor{black}{3.14} & \textcolor{black}{3.38}  \\
        & MambaDC-13 & \textcolor{black}{1.79} & \textcolor{black}{2.24} & \textcolor{black}{2.69} & \textcolor{black}{3.19} & \textcolor{black}{3.42}  \\
        \toprule[1pt]
    \end{tabular}
    \label{tab7}
\end{table}


\begin{table}[!b]
\centering
    \scriptsize
    \def\arraystretch{0.98}
    \setlength{\tabcolsep}{0.9pt}
\caption{\textcolor{blue}{Comparison to the SoTA baseline systems on the VB-DMD benchmark dataset. The best scores are in boldface. $\dagger$ denotes the results reproduced using the source code provided by the authors.}}
\scalebox{1.18}{
\begin{tabular}{llccccc }
\toprule
\textbf{Method}  & \# \textbf{Param.} & \textbf{PESQ} & \textbf{STOI}& \textbf{CSIG} & \textbf{CBAK} & \textbf{COVL} \\
\hline 
Noisy & - & 1.97 & 0.92 & 3.35 & 2.44 & 2.63 \\
\hline 
\hline
SEGAN \cite{SEGAN} & 43.2M       
& 2.16 & 0.93 & 3.48 & 2.94 & 2.80  \\
\textcolor{black}{DSEGAN \cite{DSEGAN}} & --       
& 2.35 & 0.93 & 3.55 & 3.10 & 2.93  \\
MetricGAN \cite{metricgan} & 1.89M   
& 2.86 & -- & 3.99 & 3.18 & 3.42  \\
PHASEN \cite{phasen} & 8.41M   
& 2.99 & -- & 4.21 & 3.55 & 3.62  \\
TFT-Net \cite{tftnet} & --   
& 2.75 & -- & 3.93 & 3.44 & 3.34  \\
T-GSA~\cite{tgsa} & --   
& 3.06 & -- & 4.18 & 3.59 & 3.62  \\
DEMUCS \cite{demcus} & 58M  
& 3.07 & 0.95 & 4.31 & 3.40 & 3.63  \\
\textcolor{black}{MHSA+SPK \cite{koizumi2020speech}} & -- 
& 2.99 & -- & 4.15 & 3.42 & 3.57 \\
\textcolor{black}{HiFi-GAN \cite{hifigan}} & -- 
& 2.94 & -- & 4.07 & 3.07 & 3.49 \\
\textcolor{black}{WaveCRN \cite{wavecrn}} & 4.66M   
& 2.64 & -- & 3.94 & 3.37 & 3.29  \\
\textcolor{black}{DCCRN \cite{dccrn}} & 3.7M   
& 2.68 & 0.94 & 3.88 & 3.18 & 3.27  \\
DCCRN+ \cite{dccrn+}& 3.3M   
& 2.84 & -- & -- & -- & --  \\
\textcolor{black}{S-DCCRN \cite{sdccrn}} & 2.34M   
& 2.84 & 0.94 & 4.03 & 2.97 & 3.43  \\
\textcolor{black}{SADNUNet \cite{sadnunet}} & 2.63M 
& 2.82 & 0.95 & 4.18 & 3.47 & 3.51 \\
\textcolor{black}{DCTCN \cite{dctcn}} & 9.7M 
& 2.83 & -- & 3.91 & 3.37 & 3.37 \\
\textcolor{black}{CleanUNet \cite{cleanunet}} & 46.07M 
& 2.91 & \textbf{0.96} & 4.34 & 3.42 & 3.65 \\
SA-TCN~\cite{sa-tcn} & 3.76M   
& 2.99 & 0.94 & 4.25 & 3.45 & 3.62  \\
DeepMMSE~\cite{DeepMMSE} & 1.98M   
& 2.95 & 0.94 & 4.28 & 3.46 & 3.64  \\
SE-Conformer~\cite{conformer-se} & --   
& 3.13 & 0.95 & 4.45 & 3.55 & 3.82  \\
MetricGAN+ \cite{metricgan+} & --   
& 3.15 & -- & 4.14 & 3.16 & 3.64  \\
SEAMNET\cite{seamnet} & 5.1M   
& -- & -- & 3.87 & 3.16 & 3.23  \\
SGMSE+~\cite{segmse} & --   
& 2.96 & -- & -- & -- & --  \\
StoRM~\cite{storm} & 27.8M   
& 2.93 & -- & -- & -- & --  \\
SGMSE+M~\cite{storm} & --   
& 2.96 & -- & -- & -- & --  \\
ResTCN+TFA-Xi~\cite{tfaj} & 1.98M   
& 3.02 & 0.94 & 4.32 & 3.52 & 3.68  \\
\textcolor{blue}{DNSIP~\cite{dnsip}} & \textcolor{blue}{1.59M}   
& \textcolor{blue}{3.17} & \textcolor{blue}{0.95} & \textcolor{blue}{4.27} & \textcolor{blue}{3.64} & \textcolor{blue}{3.74}  \\
\textcolor{black}{MP-SENet-Conformer}$^\dagger$~\cite{mpsenet} & 2.05M   
& \textcolor{black}{3.33}& \textcolor{black}{\bf 0.96}& \textcolor{black}{4.52}& \textcolor{black}{3.78}& \textcolor{black}{3.98}  \\
\textcolor{black}{MP-SENet-BiMambaDC-4} & 1.69M   
& \textcolor{black}{\bf 3.39}& \textcolor{black}{\bf 0.96}& \textcolor{black}{\bf 4.61}& \textcolor{black}{\bf 3.79}& \textcolor{black}{\bf 4.08}  \\
\textcolor{black}{MP-SENet-BiMambaDC-9} & 2.36M   
& \textcolor{black}{\bf 3.48}& \textcolor{black}{\bf 0.96}& \textcolor{black}{\bf 4.69}& \textcolor{black}{\bf 3.87}& \textcolor{black}{\bf 4.17}  \\
\toprule
\end{tabular}}
\label{demand}
\vspace{-0.5em}
\end{table}

\begin{figure*}[h]
\begin{center}
\centerline{\includegraphics[width=0.99\textwidth]{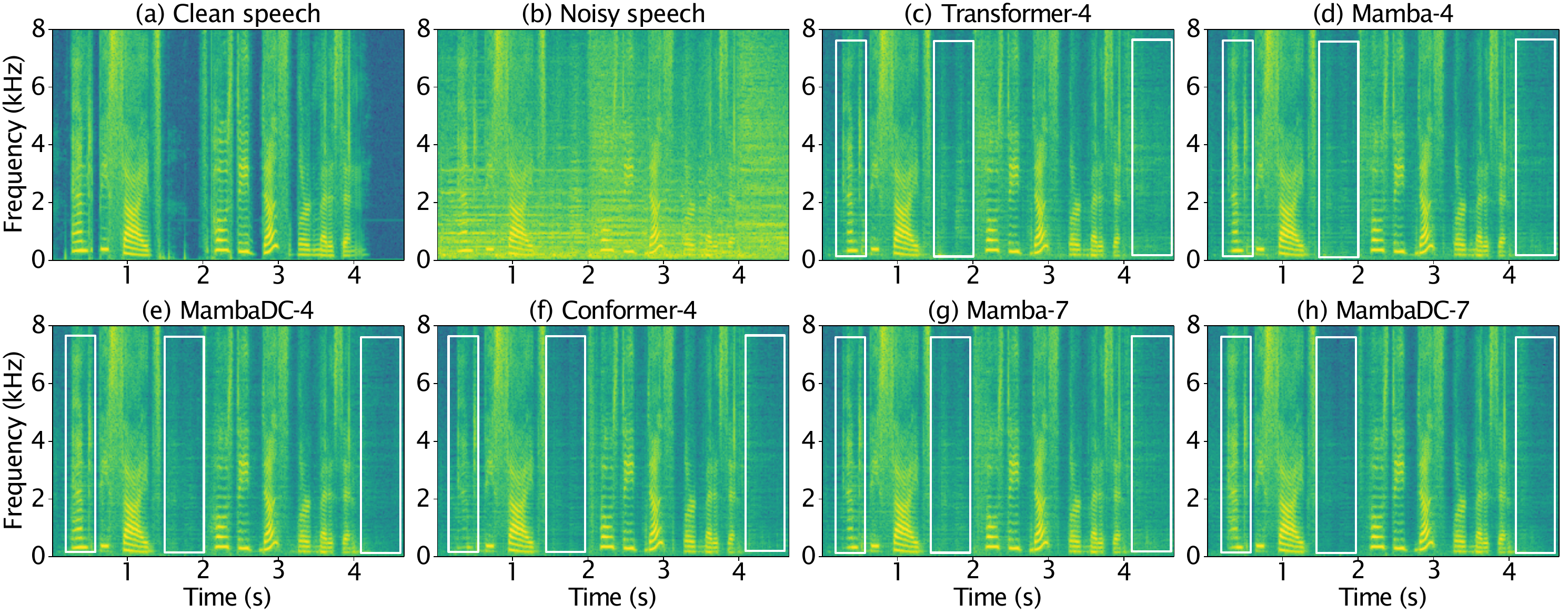}}
\caption{\textcolor{black}{Illustration of magnitude (log scale) spectrograms of clean speech, noisy speech, and enhanced speech generated by Transformer-4, Mamba-4, MambaDC-4, Conformer-4, Mamba-7, and MambaDC-7.}}
\label{fig_spectrum}
\end{center}
\end{figure*}

\textcolor{black}{In Tables \ref{tab5}--\ref{tab7}, we present the performance scores of different models in terms of CSIG, CBAK, and COVL (averaged across four noise conditions), respectively. The best results are denoted in boldface numbers. It can be clearly observed that MambaDC always provides substantial performance improvements to Mamba in all the three metrics, across IRM and PSM. In the case of $5$ dB SNR (on PSM), the MambaDC-4 and MambaDC-7 respectively improve Mamba-4 and Mamba-7 by $0.16$ and $0.19$ in CSIG, $0.12$ and $.12$ in CBAK, and $0.16$ and $0.20$ in COVL. In addition, similar to the results in Tables \ref{tab1}--\ref{tab4}, MambaDC substantially outperforms Transformer and Conformer in all the three metrics. Taking the case of $0$ dB (on IRM), MambaDC-7 and MambaDC-13 outperforms Transformer and Conformer by $0.27$ and $0.20$ in CSIG, $0.16$ and $0.12$ in CBAK, and $0.23$ and $0.18$, respectively.}

\textcolor{black}{Fig.~\ref{fig_spectrum} displays an example of spectrograms of clean speech, noisy speech, and enhanced speech produced by different models. We can observe that enhanced speech by MambaDC exhibits better noise suppression performance with less speech distortion compared to the enhanced speech by Transformer, Conformer, and the original Mamba models.}

\subsection{\textcolor{black}{Comparative Study}}
\textcolor{black}{\textcolor{black}{In Table~\ref{demand}, we present the model comparison in performance on the VB-DMD benchmark dataset.} MP-SENet~\cite{mpsenet}, a Conformer-based speech enhancement system, demonstrates state-of-the-art (SoTA) results. To further validate the superiority of our MambaDC, built upon MP-SENet model, we replace the noncausal Conformer layers with the bidirectional MambaDC (BiMambaDC) layers. We reimplemented MP-SENet using the source code provided by the authors, maintaining the same configurations. All the models are trained for $150$ epochs. It can be observed that MP-SENet with BiMambaDC layer demonstrates better performance than MP-SENet, while using fewer parameters. In addition, Mamba-based MP-SENet has a lower \textcolor{black}{memory requirement} than the original MP-SENet with Conformer. We can train the Mamba-based MP-SENet model on a single V100 GPU, whereas the original MP-SENet suffers from the out-of-memory (OOM) issue.}

\section{Conclusion}\label{sec:7}
\textcolor{black}{In this paper, we investigate the recent advanced selective state space model, i.e., Mamba, for monaural speech enhancement. To be specific, we propose using the depth-wise convolution to enhance the learning of local information for Mamba architecture, resulting in our proposal, MambaDC. Our experiments study the superiority of the MambaDC design over Mamba, across different model sizes and two commonly used training objectives, i.e., IRM and PSM. The experimental results demonstrate that our MambaDC outperforms Mamba by a large margin, in all the five measures (i.e., PESQ, ESTOI, CSIG, CBAK, and COVL). In addition, the comparison results of the MambaDC to recent state-of-the-art (SoTA) network architectures, Transformer and Conformer further demonstrate the superiority of MambaDC in performance and parameter efficiency. In our future study, we will further explore the MambaDC for audio representation learning, audio visual learning such as audio visual pre-training, as well as other speech related applications.}

\ifCLASSOPTIONcaptionsoff
  \newpage
\fi



\bibliographystyle{IEEEtran}
\bibliography{IEEEabrv,./myreference}
%

%
\vspace{-2em}
\end{document}